\documentclass{iopart}
\usepackage{graphicx}
\usepackage{iopams}
\begin{document}

\title{Atomic lattice excitons: from condensates to crystals}
\author{A. Kantian$^{1,2}$, A. J. Daley$^{1,2}$, P. T\"orm\"a$^{3}$, P. Zoller$^{1,2}$}
\address{$^1$ Institute for Quantum Optics and Quantum
Information of the Austrian Academy of Sciences, A-6020 Innsbruck,
Austria}
\address{$^2$ Institute for Theoretical Physics, University of
Innsbruck, A-6020 Innsbruck, Austria}
\address{$^3$ Nanoscience Center,
Department of Physics, P.O. Box 35, FIN-40014 University of
Jyv\"askyl\"a, Finland}
\ead{Adrian.Kantian@uibk.ac.at}

\begin{abstract}
We discuss atomic lattice excitons (ALEs), bound particle-hole pairs
formed by fermionic atoms in two bands of an optical lattice. Such a
system provides a clean setup to study fundamental properties of
excitons, ranging from condensation to exciton crystals (which appear for a large
effective mass ratio between particles and holes). Using both
mean-field treatments and 1D numerical computation, we discuss the
properities of ALEs under varying conditions, and discuss in
particular their preparation and measurement.
\end{abstract}

\section{Introduction}
\label{section:intro}

Excitons, the bound state of a fermionic particle and a fermionic
hole, belong to the most basic excitations in any system of
fermionic particles confined in the band structure of a periodic
potential~\cite{KeldyshKozlov,MoskalenkoSnoke}. In a semiconductor,
metastable excitons are created by optically pumping electrons from
the valence to the conduction band, and Coulomb repulsion between
the electrons then leads to the formation of a bound state of the
electron with the hole it left behind. At sufficiently low
temperatures, the approximately bosonic statistics of the composite
objects give rise to a superfluid state corresponding to a
condensate of
excitons~\cite{KeldyshKozlov,MoskalenkoSnoke,HanamuraHaug,GriffinStringari,KasprzakDang,Snoke,BaliliWest,LevitovButov,CombescotCombescot,CombescotDubin}.

In the present work we investigate the properties of bound
particle-hole pairs of fermionic atoms confined in an optical
lattice, where two-particle interactions are provided by collisional
processes between atoms
\cite{ChinKetterle,RomBloch,KohlEsslinger,OspelkausBongs}, or Atomic
Lattice Excitons (ALEs)~\cite{Lee}. Optical Lattice
systems~\cite{OspelkausBongs,JakschZoller,WinklerZoller,BlochZwerger,LewensteinSen,MandelBloch1,StoferleEsslinger,FallaniInguscio,SpielmanPorto},
have emerged as a very attractive tool for the study of many-body
lattice models (see,
e.g.,~\cite{SanperaLewenstein,MicheliZoller,ScarolaDasSarma,LiuWu,DuanLukin,Garcia-RipollCirac,SantosLewenstein}
particularly because of the high degree of available control over
system parameters (e.g., control over interactions using Feshbach
resonances
~\cite{KoehlerJulienne,DonleyWieman,LoftusJin,TheisHeckerDenschlag}.
These systems also allow for a wide range of measurement techniques,
giving access to quantities including momentum and quasimomentum
distribution, and higher-order correlation functions via noise
correlation measurements~\cite{AltmanLukin,FollingBloch,GreinerJin}.

ALEs constitute a novel realisation of excitons, allowing
fundamental aspects of excitons as composite objects and as an
interacting many body system to be investigated in a particularly
clean and controlled setting. In fact, the many body physics
represented by ALEs corresponds closely to the basic theoretical
models, which have been developed in seminal work in the context of
exciton condensates in semiconductors physics, focusing mainly on
many body physics of excitons as interacting composite objects of
two fermionic electron-hole constituents. A semiconductor
environment constitutes, of course, a much more complex system with
coupling to other degrees of freedom, e.g., lattice phonons and
vacuum modes of the radiation field, which are essentially absent in
atomic lattices. Furthermore, additional parameter regimes can be
accessed for ALEs, producing quantum phases that are not normally
observed in semiconductor systems. This includes, for example, the
formation of exciton crystalline structures in regimes where the
effective mass of particles and holes on the lattice are
substantially different.

\begin{figure}[tbp]
\includegraphics[width=8.5cm]{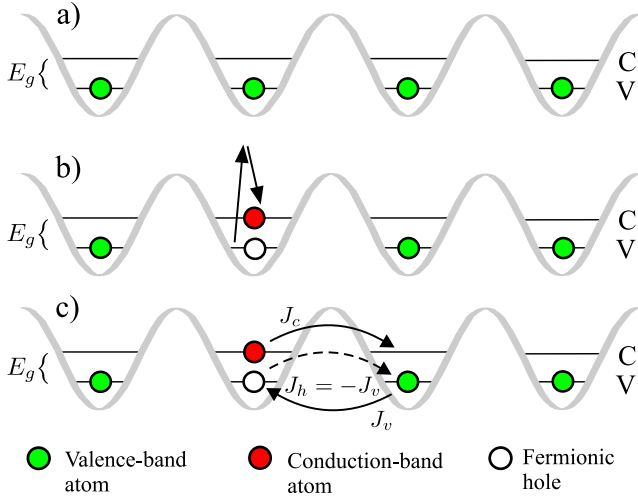}
\caption{Atomic Lattice Excitons (ALEs): (a) ALEs can be prepared
from a fermionic band insulator, with one atom per site in the
valence band (denoted V). (b) Pairs of conduction band atoms and
valence band holes are created by transferring valence band atoms to
the conduction band (denoted C), which is separated from the valence
band by the band gap energy $E_g$. c) Particle-hole pairs move by
tunnelling of the conduction-band atom with rate $J_c$ and back
tunnelling of a valence-band atom with rate $J_v$, which corresponds
to co-tunnelling of the hole. Co-tunnelling occurs because of the
interband atom-atom repulsion $U$. Thus, atom-hole pairs move
together, forming the exciton.}\label{fig:latex}
\end{figure}

In the simplest case, ALEs can be formed in a system of
spin-polarised fermions in a optical lattice, where the lowest Bloch
band (called the \emph{valence} band) is initially fully occupied
(figure~\ref{fig:latex}a). This can be set up in 1D, 2D or 3D,
although in this article we will primarily concentrate on the 1D
case. This is both the conceptually simplest case as only a single
excited Bloch band needs to be considered, and also provides
interesting analogies to the possible formation of excitons in
quantum wires. The lattice is assumed to be sufficiently deep for a
tight-binding approximation~\cite{JakschZoller} to be valid for the
first two Bloch bands, which are separated by the band gap $E_{g}$.
Atomic lattice excitons (ALEs) are created from the ground state by
exciting atoms with a laser driven Raman process into the next
highest Bloch band (called the \emph{conduction} band), where they
have a tunnelling rate $J_{c}$. The excitation leaves a fermionic
hole in the valence band, (figure~\ref{fig:latex}b), moving with
hopping rate $J_{h}=-J_{v}$, where $J_{v}$ is the valence band
hopping rate of the atoms. While spin-polarised fermionic atoms
have, by symmetry, no s-wave scattering, a repulsive collisional
interaction $U$ can be generated between conduction and valence-band
atoms in the same lattice site (e.g., by using an off-resonant Raman
transition to mix in some component of a different spin state for
particles in one of the two bands.). This on-site repulsion, in
turn, gives rise to an effective attraction between conduction-band
atoms and holes.

In this work we investigate ALEs, applying both mean-field theory
and exact numerical calculations in 1D. We derive the Fermi-Hubbard
Model description of ALEs, and then  discuss both single ALEs
(Sec.~\ref{section:prop}A) and interaction between two ALEs
(Sec.~\ref{section:prop}B). In particular, interactions between
long-lived ALEs, which due to the large effective mass ratio between
particles and holes can be analysed in a Born-Oppenheimer
approximation, are characterised by a finite range repulsion. In
Sec.~\ref{section:manybody} we then discuss the resulting many-body
phases, including ALE condensates, and a crystal phase that arises
from the effective long-range interactions. We discuss how the
properties of these phases can be measured using RF Spectroscopy and
noise correlation techniques (Sec.~\ref{section:spec}), and discuss
schemes to prepare well-defined filling factors of excitons in a low
energy many-body state in an optical lattice
(Sec.~\ref{section:formation}). We give a summary in
Sec.~\ref{section:summary}. In~\ref{sec:app1} we provide extra
details of our solution for a single exciton on the lattice, while
in~\ref{app:Nimpurities} we detail the solution to the problem of
$N$ static lattice impurities that underlie our Born-Oppenheimer
approximation. Finally, in~\ref{sec:appA} we provide additional
information about the general treatment of excition condensation.

\section{Atomic lattice excitons (ALEs)}\label{section:prop}

We consider a gas of ALEs, formed in an optical lattice that is
sufficiently deep for a tight-binding
approximation~\cite{JakschZoller} to be valid for the first two
Bloch bands, which are separated by the band gap $E_g$.  If the
lattice is isotropic or near-isotropic, there will be degenerate
lowest excited p-bands, as discussed in \cite{LiuWu}. In this paper
we focus on excitons that are tightly confined in two dimensions so
that we consider excitons in 1D, and will also give mean-field
results that are applicable to excitons in 2D or 3D with strong
anisotropies so that the p-band degeneracy is lifted, and one of the
bands can be chosen as the conduction band. In the conduction band,
atoms have a tunnelling rate $J^{\textbf{xx}'}_c$ between
neighbouring sites with lattice vectors $\textbf{x}$ and
$\textbf{x}'$ (these tunnelling rates can, in general, be
anisotropic). The excitation then leaves a fermionic hole in the
valence band, (figure~\ref{fig:latex}b), moving with hopping rate
$J^{\textbf{xx}'}_h=-J^{\textbf{xx}'}_v$, where $J^{\textbf{xx}'}_v$
is the valence band hopping rate of the atoms between neighbouring
sites.  It is convenient to expand the field operators of the
lattice-confined fermions in the localised Wannier modes
$w^{\{c,v\}}(\textbf{r}-\textbf{x})$ and the corresponding
annihilation operators $c^{\{c,v\}}_{{\bf x}}$ of conduction and
valence band respectively~\cite{JakschZoller}
\begin{equation}\psi^{\{c,v\}}(\textbf{r})=\sum_{{\bf x}} c^{\{c,v\}}_{{\bf x}} w^{\{c,v\}}(\textbf{r}-\textbf{x}),\label{eq:wanniermodes}\end{equation}
where ${\bf x}= a (x_1,\ldots,x_D)$, $(x_1,\ldots,x_D)\in\mathbb{Z}^D$ and $a$ is the
lattice spacing.

A repulsive interaction $U$ between conduction and valence-band
atoms in the same lattice site can be generated as discussed in sec.
\ref{section:intro}, which then gives rise to an effective
attraction between conduction-band atoms and holes. A Hamiltonian
describing the interacting populations of atoms in two bands can be
obtained by inserting the decomposition \ref{eq:wanniermodes} into
the two-species Hamiltonian in second quantization. We obtain a
repulsive two-species Hubbard-Hamiltonian,
\begin{equation}H=-\sum_{\langle \textbf{xx}'\rangle} J^{\textbf{xx}'}_c c^{\dagger}_{{\bf x}} c_{{\bf x}'} -
\sum_{\langle \textbf{yy}'\rangle} J_v^{\textbf{yy}'} b^{\dagger}_{{\bf y}} b_{{\bf y}'} + U \sum_{{\bf x}}
c^{\dagger}_{{\bf x}} b^{\dagger}_{{\bf x}} b_{{\bf x}} c_{{\bf x}}
\end{equation}
where $c_{{\bf x}}^{\dagger}$ and $b_{{\bf x}}^{\dagger}$ are atom
creation operators on lattice site ${\bf x}$ in conduction and
valence-band respectively, obetying the usual fermionic commutation
relations. This Hamiltonian is valid in the limit where $U,J_c,J_v
\ll E_g$, where $E_g$ is the band gap. This is well fulfilled by
typical experimental parameters. Note that terms involving transfer
of particles from one band to another do not appear because of this
condition, and therefore the total number of valence band atoms and
the total number of conduction band atoms are conserved separately.

Holes appear through the introduction of fermionic hole
creation operators $d_{{\bf x}}^{\dagger}$ in the lower band by the
definition $d_{{\bf x}}^{\dagger}=b_{{\bf x}}$.
This trivial substitution yields the particle-hole Hamiltonian
\begin{equation}\label{ham}H=-\sum_{\langle xx'\rangle} J^{\textbf{xx}'}_p c^{\dagger}_{{\bf x}} c_{{\bf x}'} -
\sum_{\langle \textbf{yy}'\rangle} J^{\textbf{xx}'}_h d^{\dagger}_{{\bf y}} d_{{\bf y}'} - U \sum_{{\bf x}}
c^{\dagger}_{{\bf x}}d^{\dagger}_{{\bf x}} d_{{\bf x}} c_{{\bf x}}
\end{equation}
where $J^{\textbf{xx}'}_p=J^{\textbf{xx}'}_c$,
$J^{\textbf{xx}'}_h=-J^{\textbf{xx}'}_v$ are now the hopping rates
of conduction band particles and holes respectively, and $U$ is the
strength of the particle-hole attraction. for conduction band atoms
and valence band holes, respectively. The attractive interaction may
give rise to bound states of conduction band atoms and valence band
holes, i.e., ALEs. Although we do not consider this case here, note
that $U<0$ would give rise to repulsively bound excitons, in the
same sense as repusively bound atom pairs, which were recently
observed experimentally~\cite{WinklerZoller} (These would consist of
bound pairs where the bound state is higher in energy than the
continuum of scattering states, instead of the stable bound states
where the bound state has a lower energy than the scattering
continuum). In the case of excitons confined to move in 1D, we use
the notation $J_p=J_p^{\textbf{xx}'}$ and $J_h=J_h^{\textbf{xx}'}$.
In the setup we consider, we will have $J_h<0$ and $J_p<0$.

\subsection{The single ALE}\label{ssection:singale}
In order to obtain the wavefunction for a single ALE, we make an ansatz for the exciton creation
operator:
\begin{equation}\hat{A}^{\dagger}|v\rangle=\sum_{\textbf{xy}} \phi_{{\bf x},{\bf y}} c^{\dagger}_{{\bf x}} d^{\dagger}_{{\bf y}}|v\rangle,\end{equation}
where $|v\rangle$ is the particle-hole vacuum, and $\phi_{{\bf
x},{\bf y}}$ is determined from solution of the eigenvalue equation
$H\hat{A}^{\dagger}|v\rangle=E\hat{A}^{\dagger}|v\rangle$. We will
now investigate the form of $\phi_{x,y}$ for the case of excitons
confined in 1D, a result which can also be straightforwardly
generalised to 2D or 3D \cite{WinklerZoller,orso2body}.

We can obtain a simple analytical solution that has the same form
either for the case $J_p=J_h$, or the case of strongly imbalanced
hopping rates, $|J_p| \gg |J_h|$. In the first case, introducing
Center-of-mass (COM) and relative coordinates ${R}=({x}+{ y})/2$,
${r}={ x}-{ y}$ gives rise to the product ansatz
\begin{equation}\phi_{{ x},{y}}=\exp\left[\rmi{K}({x}+{ y})/2\right]\rho_{{x}-{ y}},\end{equation}
which solves the eigenvalue equation exactly. The COM quasimomentum
${K}\in[-\pi/a,\pi/a]$ is factored out, and we are left with
equations for the relative wavefunction $\rho_{{r}}$:
\begin{equation}
\label{exeq}-J(\rho_{{r}+1}+\rho_{{r}-1})-U\delta_{{r}{0}}\rho_{{r}}=E\rho_{{r}}.
\end{equation}
Here $J \equiv J_{K}=2J_p\cos(Ka/2)$ is the effective hopping rate
in relative coordinates and $\delta_{{r}{0}}$ the Kronecker-delta.
Note that the relative wavefunction $\rho_{{r}}$ still depends on
the centre-of-mass quasimomentum ${K}$.

The parameters encountered in an experiment for the suggested
implementations of ALEs will typically fulfill $|J_h|/|J_p|\ll 1$,
so that holes will appear much heavier than the conduction band
atoms. We can make the approximation that the centre of mass is
approximately located at the position of the hole, and obtain in
this sense a Born-Oppenheimer wavefunction for the single exciton,
\begin{equation}\phi_{{ x},{y}}=C(y)\rho_{{x}-{ y}},\end{equation}
where $C(y)$ is the wavefunction for the hole. This leads
to~\ref{exeq}, but now with $J\equiv J_p$.

\begin{figure}[bt]
\includegraphics[width=8.5cm]{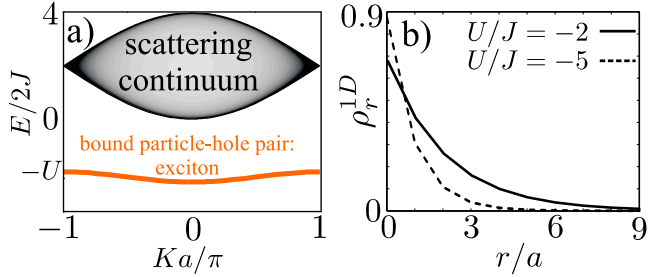}
\caption{(a) Energy eigenvalues for a single particle and a single
hole on a 1D lattice plotted as a function of COM quasimomentum, as
found from solution of~(\ref{exeq}), with $J\equiv J_K$ (see text).
The lower curve represents the Bloch Band of a single bound
atom-hole pair, the ALE. The upper portion of the spectrum is the
scattering continuum of the unbound atom and hole, where the shading
corresponds to the density of states (darker shading for higher
density of states). (b) Bound-state relative wavefunctions at
different binding energies $E_b/J$, valid both for $J_h=J_p$
($J\equiv J_K$, see text), and also in the limit $|J_h|\ll
|J_p|\equiv J$. Weakly bound states have an appreciable range over
10 or 20 sites. This falls substantially for moderately bound states
(dashed line).} \label{fig:exdisp}
\end{figure}

Eq.~\ref{exeq} can be solved exactly either by direct solution of
the difference equation in 1D using a standard exponential ansatz,
or using Green's function methods (see~\ref{sec:app1}). For the case
of $J_p=J_h$, It yields a single bound state solution for each value
of the COM quasimomentum $K$, \footnote{Note that whilst bound
states exist in 1D for any positive value of $U$, if these equations
are solved in 3D, then a bound state only appears above a critical
value for the interaction strength $(U/|J)_{crit}\approx3.95$.}, and
a continuum of solutions describing unbound states, illustrated in
figure~\ref{fig:exdisp}a. The continuum solutions correspond to
scattering states, whilst the bound state solution appears as a
Bloch band for the composite object, i.e., the ALE. The wavefunction
$\rho_{{r}}$ for the bound state solution is given by
\begin{equation}\label{singlealewf}\rho_{{r}}=C \nu ^ {-
|r/a|},\quad \nu = \frac{-E-\sqrt{(E+2J)(E-2J)}}{2J}\end{equation}
\begin{equation}\label{dispersionequation}E=-\sqrt{U^2+(2J)^2} \end{equation}
where $C$ is the normalisation constant and $E<0$ the energy of the
bound state solution. This is depicted in figure~\ref{fig:exdisp}b.
As $U/|J|$ increases, the ALE becomes more tightly bound, and
$\rho_r$ decays more rapidly. For $J>0$, the phase of the
wavefunction is constant (given by the phase of $C$), whereas for
$J<0$, the sign of the $\rho_r$ will oscillate between neighbouring
sites.

\subsection{Tightly Bound ALEs}
\label{sssection:tightaleinter}

In the tightly bound limit, $|J_p|,|J_h| \ll U$, we can interpret
ALEs as effective hard-core bosons with creation operators
$b_{{x}}^{\dagger}=c^{\dagger}_{{ x}} d^{\dagger}_{{ x}}$
\cite{MicnasRobaszkiewick}. The effective Hamiltonian is given by
\begin{equation}
H_{HC}=-J_{\rm eff}\sum_{\langle i j \rangle} b_i^\dag b_j +  V_{\rm eff} \sum_{\langle i j \rangle} \hat n_i \hat n_j,
\end{equation}
where $\hat n_i=b_i^\dag b_j$. The effective exciton hopping rate
$J_{\rm eff}$ can be calculated in degenerate second-order
perturbation theory \cite{MicnasRobaszkiewick}, and the
dimension-independent result is $J_{\rm eff}=2J_p J_h/U$. A weak
nearest-neighbour repulsion also appears in second-order
perturbation theory, with $V_{\rm eff}=(J_p^2+J_h^2)/U$ again
independent of dimensionality~\cite{Lee,CazalillaGiamarchi}. Note
that in this limit, where the atoms are treated as hard-core bosons,
this form for the interaction precludes the formation of bound
states (i.e., bi-excitons).

\subsection{Interaction of ALEs: The Born-Oppenheimer approximation}\label{ssection:aleinter}
In the typical experimental situation, where $J_p \gg J_h$, we can
treat ALE-ALE interactions in a Born-Oppenheimer (BO) approximation,
and observe longer range interactions mediated by the faster
tunnelling conduction band atoms. We find it convenient to derive
the results in this limit for the general case in $D$ dimensions,
although we will again primarily apply them in the 1D case.

The basic idea is that we initially assume the holes to be
essentially static (i.e. $J_h\approx 0$) at a given set of positions
$\underline{\textbf{R}}=\{\textbf{R}_1,\textbf{R}_2,\ldots,\textbf{R}_N\}$.
The conduction band atoms then move with hopping rate $J_p$ in the
static potential given by the holes, and have coordinates denoted
$\underline{\textbf{r}}=\{\textbf{r}_1,\textbf{r}_2,\ldots,\textbf{r}_N\}$.
In this sense, we begin by decomposing the full Hamiltonian into two
parts, one describing slow-moving holes ($H_h$) and one particle
motion and particle-hole interaction $H_{BO}$. In first quantisation
we obtain
\begin{eqnarray}H&=&H_h+H_{BO}\nonumber\\
H_h&=&-\sum_{n=1}^N J_h\widetilde{\Delta}_{\textbf{R}_n}\\
H_{BO}&=&\sum_{n=1}^N \left[-J_p\widetilde{\Delta}_{\textbf{r}_n}
-U\sum_{n'=1}^N\delta_{\textbf{r}_n\textbf{R}_{n'}}\right],\label{HBOequation}\end{eqnarray}
where the operator
\begin{equation}
\tilde{\Delta}_{{\bf x}} \Psi({\bf x},{\bf y})\!=\! \sum^{D}_{d=1}\left[
\Psi({\bf x\!+\!e}_{d},{\bf y}) \!+\! \Psi({\bf x\!-\!e}_{d},{\bf y}) - 2
\Psi({\bf x},{\bf y})\right]
\end{equation}
denotes a discrete lattice Laplacian on a cubic lattice with unit
vectors ${\bf e}_{d}$ in $D$ dimensions. Note that we have shifted
the zero of energy by $N(2J_p+2J_h)$ for convenience in writing the
discrete Laplacian.

We write the full time-dependent many-body wavefunction of the ALEs as
\begin{equation}\label{BOwavedecompose}\psi(\underline{\textbf{r}},\underline{\textbf{R}},t)=\sum_{\alpha}
C_{\alpha}(\underline{\textbf{R}},t)\phi_{\alpha}(\underline{\textbf{r}};\underline{\textbf{R}}).\end{equation}
Here, the functions
$\phi_{\alpha}(\underline{\textbf{r}};\underline{\textbf{R}})$ are
solutions of the Schr\"odinger equation for motion of the particles
in the static potential provided by the holes,
\begin{equation}\label{BOeigeneq1}H_{BO}[\underline{\textbf{R}}]\phi_{\alpha}(\underline{\textbf{r}};\underline{\textbf{R}})=
E_{\alpha;BO}[\underline{\textbf{R}}]\phi_{\alpha}(\underline{\textbf{r}};\underline{\textbf{R}}).
\end{equation}
where $\underline{\textbf{R}}$ are considered parameters.

In order to obtain an equation for the functions
$C_{\alpha}(\underline{\textbf{R}},t)$, we apply the full
Hamiltonian to~(\ref{BOwavedecompose}), multiply with
$\phi^*_{\beta}(\underline{\textbf{r}};\underline{\textbf{R}})$ and
trace over the $\underline{\textbf{r}}$s:


\begin{eqnarray}\label{BOeffectiveeq}\fl\rmi \partial_t C_{\alpha}(\underline{\textbf{R}},t)=
\left[-J_h\sum_{n=1}^N\widetilde{\Delta}_{\textbf{R}_n}+E_{\alpha;BO}[\underline{\textbf{R}}]\right]C_{\alpha}(\underline{\textbf{R}},t)\nonumber\\
-J_h\sum_{\beta}\sum_{n=1}^N\sum_{d=1}^D\sum_{\epsilon=\pm}
C_{\beta}(\underline{\textbf{R}}+\epsilon
\underline{\textbf{e}_d}^n,t)\left\{\sum_{\underline{\textbf{r}}}
\phi^*_{\alpha}(\underline{\textbf{r}};\underline{\textbf{R}})\left[\phi_{\beta}(\underline{\textbf{r}};\underline{\textbf{R}}+\epsilon
\underline{\textbf{e}_d}^n)
-\phi_{\beta}(\underline{\textbf{r}};\underline{\textbf{R}})\right]\right\},
\end{eqnarray}

where
$\underline{\textbf{e}_d}^n=\{0,\ldots,0,\textbf{e}_d,0,\ldots,0\}$
represent the $D$ different unit vectors on the lattice for the
$n$-th hole coordinate. This describes the dynamics of our system as
a multi-channel problem, with each $\alpha$ denoting a particular
channel, corresponding to an eigenstate found
from~(\ref{BOeigeneq1}). The first term describes the hole dynamics
for a single channel, with the holes moving in the effective
potential $E_{\alpha;BO}[\underline{\textbf{R}}]$. The second term
provides coupling between the channels, and has the general form of
another discrete Laplacian acting on
$C_{\beta}(\underline{\textbf{R}},t)$, where each hopping process
now carries $\underline{\textbf{R}}$- and $\alpha$-dependent
coefficients.

\begin{figure}[tb]
\includegraphics[width=8.5cm]{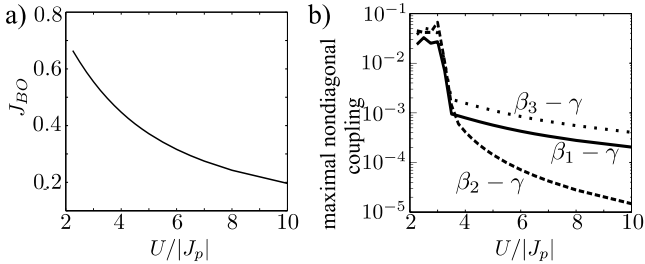}
\caption{Diagonal and nondiagonal contributions of the last term
of~(\ref{BOeffectiveeq}) for $N=3$ and $J_h/J_p=0.1$ for a $60$-site
lattice. (a) $J_{BO}$ plotted as a function of $U/|J_p|$, computed
from the diagonal elements. As ALEs become more tightly bound, the
result approaches that obtained from perturbation theory,
$J_{eff}=2J_pJ_h/U$. (b) Maximal effective coupling between the
lowest channel $\gamma$ and the first three excited channels
$\beta_1$, $\beta_2$, $\beta_3$ arising from the off-diagonal
elements,
$\max_{\underline{R},n}\{|J_p\sum_{\underline{r}}\phi_{\gamma}^*(\underline{r};\underline{R})\phi_{\beta}(\underline{r};\underline{R}\pm
\underline{e}^n)/(E_{\beta;BO}[\underline{R}\pm
\underline{e}^n]-E_{\gamma;BO}[\underline{R}])|\}$. Even for low
$U/|J_p|$, nonadiabatic coupling will always add just small
perturbative corrections, provided $|J_h/J_p|$ is sufficiently
small.} \label{fig:nonadiabatic}
\end{figure}

If the terms corresponding to coupling between the channels are
small, then we can reduce the wavefunction
expansion,~(\ref{BOwavedecompose}) to a single value of $\alpha$.
This we will do for the lowest energy solution
to~(\ref{BOeigeneq1}),
$\phi_{\gamma}(\underline{\textbf{r}};\underline{\textbf{R}})$,
which corresponds to  $N$ bound ALES. We can then interpret the
corresponding function $C_{\gamma}(\underline{\textbf{R}},t)$ as the
wavefunctions of the composite ALEs, moving in a potential obtained
from the wavefunctions
$\phi_{\gamma}(\underline{\textbf{r}};\underline{\textbf{R}})$. This
interpretation amounts to an adiabatic approximation in the ratio of
the coupling terms in~(\ref{BOeffectiveeq}) with $\gamma\neq \beta$
and the energy separation
$|E_{\gamma;BO}[\underline{\textbf{R}}]-E_{\beta;BO}[\underline{\textbf{R}}]|$
of the modes $\gamma,\beta$. This approximation will be satisfied if
the ratio $J_h/J_p \ll 1$, as the separation of the modes is at
least as large as $J_p$, see figure~\ref{fig:nonadiabatic}b for an
illustration. However, this approximation can have a much larger
range of validity in practice, if the energy difference
$|E_{\gamma;BO}[\underline{\textbf{R}}]-E_{\beta;BO}[\underline{\textbf{R}}]|$
is greater than or equal to the binding energy of a single ALE. For
example, this approximation is clearly also valid provided $J_h \ll
U$, c.f. figure~\ref{fig:nonadiabatic}b.

The diagonal terms $\alpha=\beta$ in the second rhs.-term
of~(\ref{BOeffectiveeq}) give an effective renormalisation of the
hopping rate $J_h$ that appears in the kinetic energy term in the
first line of the same equation. This assumes that the coefficients
given by the trace over $\underline{\textbf{r}}$ are approximately
equal for all $\underline{\textbf{R}}$, which we have confirmed by
evaluation of the corresponding terms from exact calculations in 1D.
We find maximum variation of the coefficients of the order of $15\%$
even when two holes are very close (within $\sim 2$ sites), whilst
for a more dilute gas, this approximation becomes even better. The
renormalised kinetic energy can be interpreted as the kinetic energy
term of the composite exciton, and in the tightly bound limit,
$|J_p|,|J_h| \ll U$, reduces to the values found in sec.
\ref{sssection:tightaleinter}.

The solution of~\ref{BOeigeneq1} is simplified in our case because
the particles are non-interacting fermions, and thus the problem
reduces to finding the solutions for a particle on a lattice with
$N$ impurities. This is detailed in~\ref{app:Nimpurities}. We assume
throughout that the $N$ impurities are sufficiently separated from
each other that $N$ bound states exist. The critical separation at
which bound states occur is dependent upon $U/|J_p|$ and the number
of particles. For $N=2$ bound states exist even when holes are on
neighbouring sites provided $U/|J_p|\gtrsim 2$.

For $N=2$, The single-particle bound states are then the symmetric
and antisymmetric superpositions $\rho_{\pm}$ (\ref{BOslater}) of
two exponentially decaying bound states, centred at $\textbf{R}_1$
and $\textbf{R}_2$. These take the same form as the single ALE
wavefunction,~(\ref{singlealewf}), but now with different energy $E$
that depends on the separation $R=|{\bf R}_1-{\bf R}_2|$. The
lowest-energy two-atom wavefunction
$\phi(\textbf{r}_1,\textbf{r}_2;\textbf{R}_1,\textbf{R}_2)$ is then
obtained from the Slater-determinant of $\rho_{\pm}$. In principle,
the choice of overall phase of these solutions can be made
differently for different $\textbf{R}_1,\textbf{R}_2$. Without loss
of generality we can choose each
$\phi(\textbf{r}_1,\textbf{r}_2;\textbf{R}_1,\textbf{R}_2)$ to be
real, leaving us with a choice as to whether
$\phi(\textbf{r}_1,\textbf{r}_2;\textbf{R}_1,\textbf{R}_2)$ should
be symmetric or antisymmetric under exchange of hole coordinates
$\textbf{R}_1$ and $\textbf{R}_2$. However, in order to minimise the
non-adiabatic terms in~(\ref{BOeffectiveeq}), we are required to
choose these to be antisymmetric under exchange of $\textbf{R}_1$
and $\textbf{R}_2$, in addition to the requirement of antisymmetry
under exchange of particle coordinates $\textbf{r}_1$ and
$\textbf{r}_2$. It then follows that in order to ensure the correct
antisymmetry for exchange of hole coordinates $\textbf{R}_i$ in the
total wavefunction
$\psi(\underline{\textbf{r}},\underline{\textbf{R}},t)$,
[(\ref{BOwavedecompose})], that
$C_\gamma(\textbf{R}_1,\textbf{R}_2,t)$ must be \emph{symmetric}
under exchange of coordinates $\textbf{R}_i$. In the limit where the
BO approximation is valid, we can then interpret the coordinates
$\textbf{R}_i$ as approximate centre of mass coordinates for the
ALEs, which behave under this approximation as composite bosons.

\begin{figure}[htb]
\includegraphics[width=7.5cm]{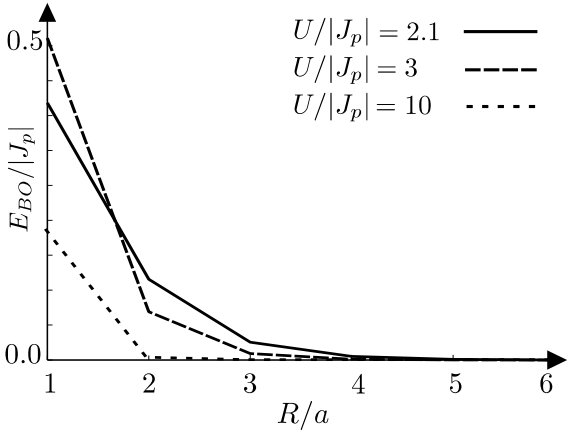}
\caption{Born-Oppenheimer curves for the interaction between two
ALEs as a function of separation $R$, for different values of $U/|J_p|$.
For $U/|J_p|\approx2-3$, longer-range interactions are significant.
For tightly bound excitons however, the Born-Oppenheimer
approximation recovers the perturbative limit, in which only
nearest-neighbour interactions are appreciable. The zero of energy is chosen so that all curves are
to zero in the limit $R/a\gg1$.} \label{fig:bopot}
\end{figure}

The lowest values of the potential $E_{\gamma;BO}[{{R}}_1,{{R}}_2]$
in 1D are shown for various values of $U/|J_p|$ in
figure~\ref{fig:bopot}, where the zero of energy is chosen so that
$E_{\alpha;BO}[R_1,R_2]\rightarrow 0$ as $R=|R_1-R_2| \rightarrow
\infty$. For large $U/|J_p|$, interactions are very short-ranged,
reducing to the values found in section
\ref{sssection:tightaleinter}. As $U/|J_p|$ is decreased, the
increase in mobility of the conduction band atoms leads first to
increased interaction strength, and then as $U/|J_p|$ is further
decreased, to longer-range interactions. These are shown in the
figure around the parameter values where they become significant
over 2-3 sites, which is approximately between $U/|J_p|=2$ and
$U/|J_p|=3$. For finite $J_h$ these interactions will be important
as long as the energy of the COM-motion is not much larger than the
interaction energy.

For $N>2$, we can always decompose the potential
$E_{\alpha;BO}[\underline{\textbf{R}}]$ into a sum over
contributions from different numbers of particles.

\begin{figure}[tb]
\includegraphics[width=8cm]{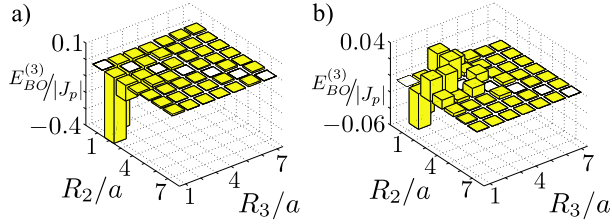}
\caption{Three-body interactions between ALEs
$E^{(3)}_{\alpha;BO}[0,{ R}_2,{ R}_3]$ computed in the
Born-Oppenheimer approximation curves for (a) $U/|J_p|=2.1$ and (b)
$U/|J_p|=3$. These results should be compared to the two-body
interactions plotted in figure~\ref{fig:bopot}. Even when the three
ALEs are close together, the three-body interactions are dominated
by two-body interactions, with three-body interactions becoming
extremely small for $U/|J_p|\gtrsim3$. This is still the case near
the limit $U/|J_p|\sim 2$, at which limit one bound state disappears
for the case of ALEs existing on neighbouring lattice sites.}
\label{fig:BO3pot}
\end{figure}

For example, for $N=3$ we can take the two-body potential
$E^{(2)}_{\alpha;BO}[{\bf R}_1,{\bf R}_2]$ found for two particles
(when the third is an infinite distance from the first two), and
define the 3-body interaction $E^{(3)}_{\alpha;BO}[{\bf R}_1,{\bf
R}_2,{\bf R}_3]$ such that
\begin{equation}
 E_{\alpha;BO}[\underline{\textbf{R}}]=E^{(3)}_{\alpha;BO}[{\bf R}_1,{\bf R}_2,{\bf R}_3] + \sum_{i<j} E^{(2)}_{\alpha;BO}[{\bf R}_i,{\bf R}_j].
\end{equation}
By solving~(\ref{BOeigeneq1}) for $N=2$ and $N=3$ we can thus assess
the importance of genuine three body interactions. Typical results
for the 1D case, in which we are primarily interested, are ploted in
figure~\ref{fig:BO3pot}. They show two typical solutions for
$E^{(3)}_{\alpha;BO}[0,{ R}_2,{ R}_3]$, and we see that only for low
$U/|J_p|$ there are any appreciable three-body interactions, even at
small distances. The same will be true for larger $N$ provided that
the density remains sufficiently small. Thus, for cold ALEs at
sufficiently small densities, the BO-potentials in
~(\ref{BOeffectiveeq}) can be decomposed into a sum of two-hole
interactions,
\begin{equation}
 E_{\alpha;BO}[\underline{\textbf{R}}]\approx \sum_{i<j} E^{(2)}_{\alpha;BO}[{\bf R}_i,{\bf R}_j].
\end{equation}

There are two cases where the symmetry properties of the two
particle wavefunction also transfer to the $N$-body case. The first
is the limit of two-body collisions, where the only significant
interactions are those between each particle and its nearest
neighbour,
\begin{equation}
 E_{\alpha;BO}[\underline{\textbf{R}}]\approx \frac{1}{2}\sum_{i} \max_j E^{(2)}_{\alpha;BO}[{\bf R}_i,{\bf R}_j].
\end{equation}
In this limit, we can construct
$\phi_{\alpha}(\underline{\textbf{r}};\underline{\textbf{R}})$ from
a Slater determinant of single particle bound states at single holes
or combinations of two holes, and the arguments presented for the
two-particle case will generalise. The other case is that in which
the holes are evenly spaced. There, the correct wavefunction is
again the Slater determinant of single particle bound states centred
at each hole, irrespective of the separation provided that the $N$
bound states exist. This problem can be seen analogously to that of
a periodic potential in free space, with the single particle bound
states at each hole playing the role of Wannier functions for this
periodic structure. In each of these cases, the optimal choice for
the Born-Oppenheimer approximation is that
$\phi_{\alpha}(\underline{\textbf{r}};\underline{\textbf{R}})$  is
antisymmetric under exchange of hole or particle coordinates, and it
follows that the wavefunction $C_\gamma(R_1,\ldots,R_N)$ will be
symmetric under exchange of coordinates $\textbf{R}_i$, as discussed
above for the $N=2$ case.

In either of these regimes, we then can write an effective
Hamiltonian for the ALEs as composite bosons with creation operators
$b_{{x}}$, which in 1D takes the form
\begin{equation}
H_{HC}=-J_{\rm BO}\sum_{\langle i j \rangle} b_i^\dag b_j +
\sum_{\langle i,l \rangle} V_l \hat n_i \hat n_{i+l},\quad V_l=
E^{(2)}_{\alpha;BO}[{ R}_0,{ R}_l],
\end{equation}
$\hat n_i=b_i^\dag b_j$, and $J_{\rm BO}$ is the renormalised
effective hopping given by the first and last terms
in~(\ref{BOeffectiveeq}). It can be approximately calculated from
the solution to the single impurity problem ($N=1$),
$\tilde\phi_{\gamma}(r_1-R_1)$, as
\begin{equation}
J_{\rm BO}\approx J_h \sum_{r_1} \tilde \phi^*_\gamma(r_1-R_1)
\tilde \phi_{\gamma}(r_1-R_1+1).
\end{equation}
In figure~\ref{fig:nonadiabatic}a, $J_{BO}$ is plotted.

In this section we have effectively shown how imbalance in the
hopping rates can enhance the strength and range of interactions
between ALEs. This will lead to the stabilization of diagonal order,
i.e. crystaline order away from half-filling, which is discussed in
detail in Sec.~\ref{ssection:alecond}. Note that this limit is
extremely relevant for ALEs, where $J_p$ will typically be an order
of magnitude larger than $J_h$, and where the lifetime of an ALE is
sufficiently long for longer range interactions to form a crystaline
structure.

\subsection{Interaction of ALEs: Numerical results}

\begin{figure}[htp]
\includegraphics[width=8.5cm]{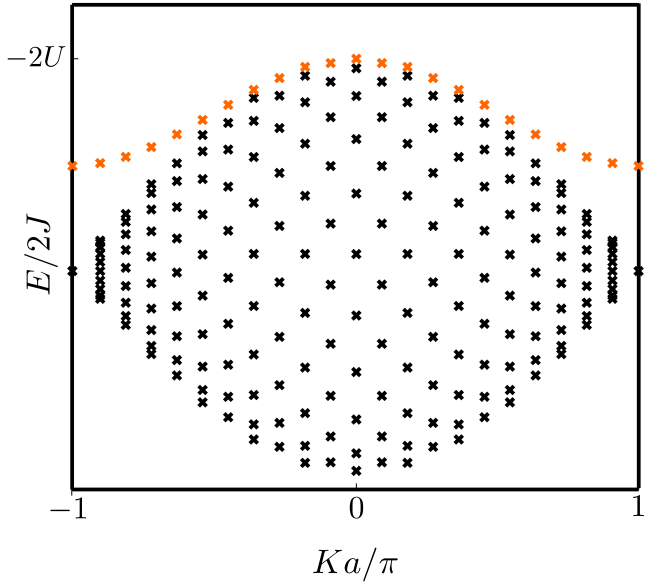}
\caption{Discrete energy eigenvalues for two interacting excitons on
22 lattice sites, plotted as a function of quasimomentum. Black
crosses correspond to scattering states. Orange crosses form the
dispersion relation of a repulsively bound state resulting from weak
exciton-exciton repulsion.} \label{fig:biexcitondispersion}
\end{figure}

Beyond the tight-binding and Born-Oppenheimer cases, we have also
performed small-scale exact diagonalizations for two interacting
excitons in 1D, using parameters ranging from $U/|J|=1$ to
$U/|J|=25$ on up to 22 lattice sites. Strong imbalances in the
hopping rates of holes and particles were considered as well,
ranging from $J_h/J_p=1$ to $J_h/J_p=1/40$. All these calculations
show only effective repulsion between the two excitons and a
complete absence of bound states of four particles. A typical result
is depicted in figure~\ref{fig:biexcitondispersion}, where the
scattering continuum and the dispersion relation of the anti-bound
state resulting from weak repulsion can be clearly distinguished.

In all cases in 1D that we have analyzed, both analytically and numerically, we only find repulsive
interactions, and thus no bound states. With increasing spatial
dimensionality, it becomes generally more unlikely for bound states
to form in scattering problems, and thus we also expect that this
result will hold for arbitrary values of $|J_h|/U,|J_p|/U$ in 2D and
3D.

\section{Many ALEs at low temperatures - exciton condensates and crystals}
\label{section:manybody}

Much fundamental interest in semiconductor excitons has stemmed from
their predicted ability to undergo condensation in 3D or to form
quasicondensates in 1D. Due to their isolation from the environment,
ALEs present an opportunity to investigate exciton condensation in a
clean and highly controllable environment. In addition, the long
lifetimes of ALEs, combined with  long range interactions that are
found in the case of unbalanced hopping give rise to other
potentially interesting phases, especially an ALE crystal.

\subsection{The ALE condensate}
\label{ssection:alecond}

Due to the effective bosonic nature of ALEs, we expect them to form
a condensate characterised by off-diagonal Long-Range Order (ODLRO)
at sufficiently low temperatures, in analogy to semiconductor
excitons \cite{MoskalenkoSnoke}. This should be true irrespective of
whether the excitons are strongly or weakly bound.

\subsubsection{Continuum approach}\label{ssection:cond}

The mean-field approximation normally made for semiconductor
excitons is a continuum model, and can be generalised to the lattice
in order to obtain a description of the ALE groundstate at
zero-temperature. In the continuum limit we write the Hamiltonian as
\begin{eqnarray}\label{scham}H&=&\sum_{{\bf k}}\left[E^p_{{\bf k}}c^{\dagger}_{{\bf k}} c_{{\bf k}}+E^h_{{\bf k}} d^{\dagger}_{{\bf k}} d_{{\bf k}}\right]
+\sum_{\textbf{klq}}\left[\frac{1}{2}V^p_{\textbf{q}}c^{\dagger}_{{\bf k}+{\bf q}}c^{\dagger}_{{\bf l}-{\bf q}}c_{\textbf{l}}c_{{\bf k}}\right.\nonumber\\
& & \left.+\frac{1}{2}V^h_{\textbf{q}}d^{\dagger}_{{\bf k}+{\bf
q}}d^{\dagger}_{{\bf l}-{\bf q}}d_{\textbf{l}}d_{{\bf k}}
+V_{\textbf{q}}c^{\dagger}_{{\bf k}+{\bf q}}d^{\dagger}_{{\bf
l}-{\bf q}}d_{\textbf{l}}c_{{\bf k}}\right],\end{eqnarray} where
$E^p_{{\bf k}}$ and $E^h_{{\bf k}}$ are the dispersion relations for
particles and holes respectively, and  $V^p_{\bf q}$, $V_{\bf q}^h$
and $V_{\bf q}$ denote the effective interaction potentials for
particle-particle, hole-hole, and particle-hole interactions,
respectively. The exciton creation operator can then be written as
\begin{equation}\label{exdef}\hat{A}^{\dagger}_{\alpha}|0\rangle=
\sum_{{\bf k},{\bf k}'}A_{\alpha}({\bf k},{\bf k}')c^{\dagger}_{{\bf
k}} d^{\dagger}_{{\bf k}'}|0\rangle,\end{equation} where the index
$\alpha$ will specify the quantum numbers for the exciton. The key
observation is that the single-exciton state of lowest energy will
have the lowest possible bound-state level $n=0$, and COM-momentum
${\bf K}=\textbf{w}$. Here, ${\bf w}$ denotes the difference in
position between the maximum of the valence band and the minimum of
the conduction band. It can either be ${\bf 0}$ or half a reciprocal
lattice vector. If ${\bf w}={\bf 0}$ (as was the case for the 1D
ALEs in the previous section) the system has a \emph{direct band
gap}. Whereas for ${\bf w}\neq {\bf 0}$ it has an \emph{indirect
band gap} (the 2D and 3D realization of ALEs are of this form).

In the following, the dispersion relations of the valencce band, is
shifted by $-{\bf w}$. Then the single exciton operator of lowest
energy with $\alpha=({\bf w},0)$ has the form
$\hat{A}^{\dagger}_{0,0}|0\rangle=\sum_{{\bf k}} A({\bf
k})c^{\dagger}_{{\bf k}} d^{\dagger}_{-{\bf k}}|0\rangle$ with
$A({\bf k})=A_{0,0}({\bf k},-{\bf k})$. Such a state corresponds to
an effective pairing of electron at ${\bf k}$ and hole at ${\bf
w}-{\bf k}$ (when the notational shift of the hole-quasimomenta is
reversed). The equivalent of this pairing in an approximate
many-body groundstate emerges from an ansatz first proposed by
Keldysh and Kozlov \cite{KeldyshKozlov}. There, the ground state of
the system is assumed to be a coherent state for this single-exciton
operator $\hat{A}^{\dagger}_{0,0}$,
\begin{equation}\label{bcsstate1}\mathcal{D}|0\rangle=\exp\left[\sqrt{n_{ex}}(\hat{A}^{\dagger}_{0,0}-\hat{A}_{0,0})\right]|0\rangle=\prod_p (u_{{\bf k}}+v_{{\bf k}}
c^{\dagger}_{\textbf{k}} d_{-{\bf
k}}^{\dagger})|0\rangle,\end{equation}
\begin{equation}\label{bcsstate2}u_{{\bf k}}=\cos(\sqrt{n_{ex}}A({\bf k}))
\quad v_{{\bf k}}=\sin(\sqrt{n_{ex}}A({\bf k})).\end{equation} The
resultant state has the same structure as the BCS ground state of
weakly correlated electrons in a superconductor. At finite density
the coefficients $A({\bf k})$ in~(\ref{bcsstate1})
and~(\ref{bcsstate2}) are no longer the single-exciton ground state
wavefunction. Instead, the coefficients $u_{{\bf k}}$ and $v_{\bf
k}$ have to be determined in a self-consistent manner
(see~\ref{sec:appA}) for background). In mean-field theory and using
the constraint of fixed density these are found to be
\begin{equation}\label{bcseq}\Delta^P_{{\bf k}}=-\sum_{\textbf{l}}V_{{\bf l}-{\bf k}}\frac{\Delta^P_{\textbf{l}}}{E^P_{\textbf{l}}},\quad
n_{ex}=\frac{1}{2}\sum_{\textbf{l}}\left(1-\frac{\xi_{\textbf{l}}}{E_{\textbf{l}}^P}\right)\end{equation}
where
\begin{eqnarray}\label{bcsxi}\xi_{{\bf k}}&=&E^c_{{\bf k}}+E^h_{{\bf k}}+2\sum_{\textbf{l}}\Sigma_{{\bf l}-{\bf k}}v_{\textbf{l}}^2\\
                \label{sigma}\Sigma_{{\bf k}}&=&V_0+\frac{1}{2}\left(V_0^p+V_0^h-V_{{\bf k}}^p
                -V_{{\bf k}}^h\right)\\
                \label{bcsdelta}\Delta^P_{{\bf k}}&=-&\sum_{\textbf{l}}V_{{\bf l}-{\bf k}}u_{\textbf{l}}v_{\textbf{l}}\\
                \label{bcsdisp}E^P_{{\bf k}}&=&\sqrt{\xi_{{\bf k}}^2+(2\Delta^P_{{\bf k}})^2}.
                \end{eqnarray}
In this regime the Hamiltonian (\ref{scham})  can be diagonalised
with new fermionic quasiparticle and quasihole operators $C_{{\bf
k}}^{\dagger}$ and $D_{{\bf k}}^{\dagger}$ [see~(\ref{newop})] to
yield
\begin{equation}H_{MF}=\sum_{{\bf k}} \left(E^{(1)}_{{\bf k}} C_{{\bf k}}^{\dagger}C_{{\bf k}}+
E^{(2)}_{{\bf k}} D^{\dagger}_{{\bf k}} D_{{\bf
k}}\right),\end{equation} with the expressions for the new
quasiparticle dispersion relations $E_{{\bf k}}^{(1)}$, $E_{{\bf
k}}^{(2)}$ given in~\ref{sec:appA},~(\ref{disp1}), (\ref{disp2}).

As a consequence of this ansatz for the groundstate, the lowest
energy solution $({\bf k},n)=(0,0)$ of the single exciton can be
recovered from the above mean-field equations in the limit of
vanishing density $n_{ex}\rightarrow 0$ ~\cite{ComteNozieres}.

The condensate thus exhibits off-diagonal long-range order, with the
associated dissipationless transport properties. For excitons,
however, these involve only dissipation-free transport of excitation
energy and momentum, not of mass or charge transfer.

\subsubsection{ALE condensate on a lattice}

We can adapt Hamiltonian~(\ref{scham}) - and the mean-field results
derived from it - directly to the case of ALEs, where $\textbf{k}$
now denotes lattice quasimomenta in the first Brillouin-zone (B.Z.)
instead of momenta in the continuum case, and we set
$V^p_{\textbf{q}}=0$, $V^h_{\textbf{q}}=0$, $V_{\textbf{q}}=-U$.
Using Hamiltonian (\ref{ham}) and applying~(\ref{bcseq}) -
(\ref{bcsdisp}), the condensate at zero temperature can be
approximated by a BCS-type groundstate
\begin{equation}|\Psi\rangle=\prod_{{\bf k}} (u_{{\bf k}}+v_{{\bf k}}
c^{\dagger}_{{\bf k}} d_{-{\bf k}}^{\dagger})|0\rangle\end{equation}
where $u_{{\bf k}}$ and $v_{{\bf k}}$ must obey the gap-equation and
the density constraint
\cite{ComteNozieres,NozieresSchmitt-Rink,FetterWalecka}:
\begin{equation}\label{alebcseq}U^{-1}=a^D\int_{B.Z.} \frac{\rmd ^Dk}{(2\pi)^D} \frac{1}{E_{\bf k}^P},\quad
n_{ex}=\frac{a^D}{2}\int_{B.Z.} \frac{\rmd ^Dk}{(2\pi)^D}
\left(1-\frac{\xi_{{\bf k}}}{E^P_{{\bf k}}}\right)\end{equation}
where
\begin{eqnarray}\label{alexi}\xi_{{\bf k}}&=&-\sum_{d=1}^3(J_p^d+J_h)\cos(k_da)-\mu_{ex}+2Un_{ex},\\
\label{aledisp}E^P_{{\bf k}}&=&\sqrt{\xi_{{\bf
k}}^2+(2\Delta^P)^2},\end{eqnarray} and $J_p^d$ denotes the hopping
rates of conduction band atom in the $d$-th direction. As the
conduction band is assumed anisotropic (see
Sec.~\ref{section:prop}), $J_p^1\neq J_p^2,J_p^3$ will hold
generally, with $J_p^1<0$, $J_p^2,J_p^3>0$~\cite{LiuWu}.
Consequently, the conduction band has its minima around $(\pm
\pi/a,0,0)$, and thus ${\bf w}=(0,\pi/a,\pi/a)$. Contrary to the
situation in 1D, the system thus has an indirect gap, and at zero
temperature the ALEs condense into a motional state with nonzero
quasimomentum.

Evaluating~(\ref{disp1}) and (\ref{disp2}) for the given model of
ALEs yields the dispersion-relations of the new quasiparticles and
quasiholes respectively:
\begin{equation}\label{aledisp1}E^{(1)}_{{\bf k}}=\frac{1}{2}\left(E_{{\bf k}}^p-E_{{\bf k}}^h+E^P_{{\bf k}}\right),
\;\; E^{(2)}_{{\bf k}}=\frac{1}{2}\left(E_{{\bf k}}^h-E_{{\bf
k}}^p+E^P_{{\bf k}}\right)\end{equation} where
\begin{equation}
E_{{\bf k}}^p=-2\sum_{d=1}^D J_p^d\cos(k_da),  \; \; E_{{\bf
k}}^h=-2\sum_{d=1}^D J_h\cos(k_da).
\end{equation}
For two-band implementation of ALEs considered here, where
$J_p^d\neq J_h$ in 2D and 3D, the quasiparticle dispersion relations
are different, $E^{(1)}_{{\bf k}}\neq E^{(2)}_{{\bf k}}$. Thus they
exhibit minima at different points in the Brillouin zone.

As shown in~\cite{ComteNozieres,NozieresSchmitt-Rink}, this
BCS-formalism can interpolate between the weak-coupling $U\ll
|J_p^d|,$ $|J_h|$ and strong-coupling $U\gg |J_p^d|,$ $|J_h|$
regime, provided the Hartree-Fock corrections to the chemical
potential are included. For strong coupling it also becomes exact in
the low-density regime ($n_{ex}\approx0$) and provides a good
qualitative description of the groundstate for increasing density up
to the case of maximal filling, $n_{ex}\approx \frac{1}{2}$.

In Sec.~\ref{section:spec} we will discuss measurements that can be
made on exciton condensates, including determination of the
condensate fraction and pairing correlations.

\subsection{The ALE crystal in 1D}\label{ssection:1dcrystal}
The long range repulsive interactions that we found for large
hopping imbalances $J_p\gg J_h$ in
Sec.~\ref{sssection:tightaleinter} suggest the possibility of
forming an ALE crystal for densities away from half filling,
$n_{ex}\neq \frac{1}{2}$. This situation is directly related to
recent research on Fermi Hubbard models with imbalanced hopping
rates for the two species \cite{CazalillaGiamarchi,DaoCapone}.

The results of the Born-Oppenheimer approximation in
Sec.~\ref{ssection:aleinter} suggest that the effective interactions
have maximal range and strength for loosely bound ALEs at
$U/|J_p|\approx 2$ - $3$ for large hopping imbalances. For these
parameters, naive inspection of figure~\ref{fig:bopot} suggests that
interaction effects will favour Diagonal Long Rangle order (DLRO),
i.e., a crystal, and suppress pairing maximally from $n_{ex}\approx
\frac{1}{3}$ upward, provided that the kinetic energy of the ALEs is
sufficiently small. We investigate the possibility of such a crystal
using imaginary time evolution on matrix product states (with a
time-depndent DMRG
algorithm)~\cite{Vidal2,Vidal1,DaleyVidal,WhiteFeiguin} to compute
the groundstate Hamiltonian (\ref{ham}) at different densities. From
the numerically calculated groundstates, we obtained
density-density- and pairing correlations for four different
densities, with $n_{ex}=\frac{15}{96}$ $\frac{23}{96}$,
$\frac{30}{96}$ and $\frac{39}{96}$, on a $96$-site lattice for
different values of $U/|J_p|$ and $J_h/J_p$.

In bosonised theories \cite{Giamarchi}, the Luttinger parameter $K$
determines the algebraic decay of correlation functions in the
thermodynamic limit at large distances (true long-range order of any
type being absent in 1D for continuously-valued fields). We expect
the pairing correlation function $\langle c^{\dagger}_x
d^{\dagger}_x d_y c_y\rangle$ to behave as $A_P/|x-y|^{K_P}$, and
the density-density-correlation function $(\langle n_x
n_y\rangle-\langle n_x\rangle\langle n_y\rangle)$ to behave as
$A_{DD}\cos(k_f(x-y))/|x-y|^{K_{DD}}$ for $|x-y|\gg 1$, where
$k_f=\pi n_{ex}$ denotes the Fermivector. $A_P$, $A_{DD}$, $K_P$ and
$K_{DD}$ were determined from fitting to these functions.

Bosonisation predicts $K_{DD}=1/K_{P}$ \cite{Giamarchi}. The
validity of bosonization will shift to decreasing wavelengths,
however, as $J_h/J_p$ decreases. This results from the increasingly
different slopes of the dispersion-relation linearisation around the
Fermi-points. Excitations that are still weak - i.e. long-wavelength
- for the more mobile conduction-band atoms, will be in the
non-linear regime for the slower atomic holes. The validity of
bosonization for any finite-size system is thus limited by the
length-scale set by the less mobile holes, and we expect our
simulation results to exhibit noticeable deviations from
bosonization predictions below some value of $J_h/J_p$.

\begin{figure}[htp]
\includegraphics[width=8.5cm]{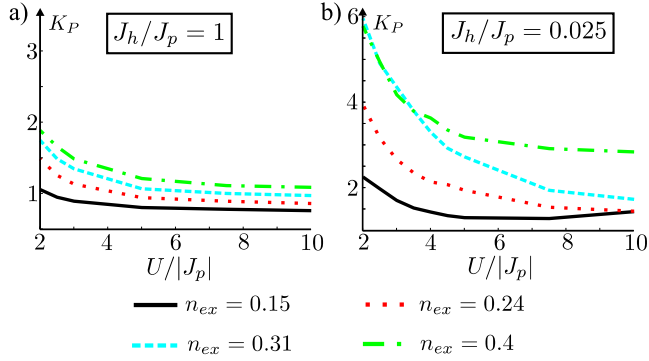}
\caption{Power of algebraic decay of pairing correlations plotted as
a function of $U/|J_p|$ for $J_p/J_h=1$ and $J_p/J_h=0.025$, and for
four different densities on a $96$-site lattice. (a) For
$J_h/J_p=1$, pairing correlations decay moderately faster with
increasing density and decreasing $U/|J_p|$. (b) For strong hopping
imbalance, pairing correlations decay much faster with decreasing
$U/|J_p|$ as ALE-ALE interaction increases in range and magnitude,
which become more significant as $n_{ex}$ increases. The decay is
especially strong for $n_{ex}\geq 0.3$ in the regime $U/|J_p|\approx
2$ $-$ $3$. From the Born-Oppenheimer theory this result is
expected, see e.g. figure~\ref{fig:bopot}, as for these parameters
and $J_h/J_p=0.025$ the next-nearest-neighbour interactions between
ALEs is still dominant over the kinetic energy of the COM-motion of
the ALEs. As a result, for $n_{ex}=0.31$, close to the ideal crystal
filling factor $n_{ex}=\frac{1}{3}$, the decay also shows the
steepest increase as $U/|J_p|$ decreases. The enhanced
next-nearest-neighbour interactions in this region strongly suppress
pairing order, and thus for the ALEs crystaline order dominates in
this regime.}\label{fig:bcscor}
\end{figure}

\begin{figure}[htp]
\includegraphics[width=8.5cm]{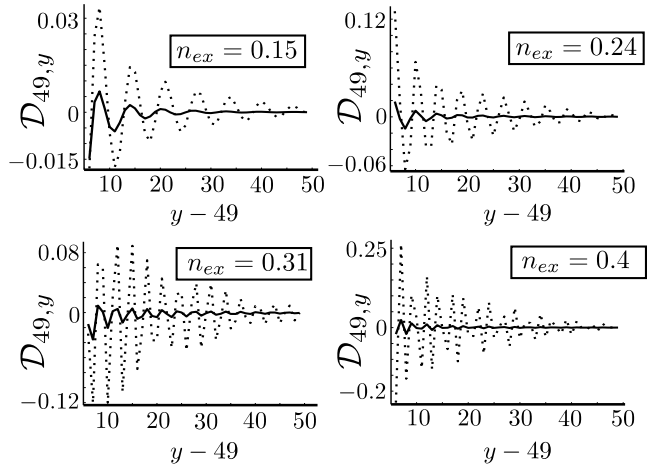}
\caption{Representative examples of real-space density-density
correlations for $J_p/J_h=1$ (solid lines) and $J_p/J_h=0.025$
(dotted lines) plotted from the center of a $96$-site lattice. Here,
$\mathcal{D}_{49,y}=\langle n_{49} n_y\rangle-\langle n_{49}\rangle
\langle n_y\rangle$ is shown at four different ALE -densities for
$U/|J_p|=3$. At all densities, both amplitude and rate of decay of
these DD-correlations are significantly improved by hopping
imbalance over the case of $J_p=J_h$. The overall magnitude of the
correlations increases with $n_{ex}$. Meanwhile, pairing
correlations start to decay rapidly in this regime of $U/|J_p|$, see
figure~\ref{fig:bcscor}. Taken together, this identifies the regime
of $|J_h/J_p|\ll 1$, $U/|J_p|=2$ $-$ $3$ and $n_{ex}\geq 0.3$ as the
region where the crystaline characteristics of the ALEs are most
pronounced.} \label{fig:ddcor}
\end{figure}

While we averaged out $k_f$-oscillations superimposed on the
algebraically decaying pairing correlations (c.f.
\cite{NoackScalapino}) to obtain reliable fits for $K_P$, this is
more difficult for $K_{DD}$ in the density-density-correlations. As
a result, we obtain large deviations from the bosonization
prediction $K_{DD}=1/K_P$ even for $J_p=J_h$, where bosonization is
most reliable. Despite this difficulty, we can identify clearly
different regimes of behaviour, dependent on the hopping imbalance
and ALE density. For illustration, the powers of algebraic decay
obtained from the fit are depicted in figure~\ref{fig:bcscor} for
the pairing correlations, while figure~\ref{fig:ddcor} shows
representative examples of density-density correlation functions in
different parameter regimes. Generally we observe in our results
that superfluid order is dominant for equal or moderately imbalanced
hopping rates, but becomes more and more suppressed as ALE density
and hopping imbalance increase. Complementary,
density-density-correlations start decaying slower and become the
dominant order. As is to be expected in 1D, this
superfluid-to-crystal transition is continuous. Crystaline behaviour
is especially pronounced in the region predicted from
Born-Oppenheimer calculations, i.e. as $n_{ex}\geq \frac{1}{3}$ and
$U/|J_p|=2$ $-$ $3$, where density-density-correlations decay only
moderately (figure~\ref{fig:ddcor}) and ALE-interactions suppress
pairing very strongly (figure~\ref{fig:bcscor}).

For higher dimensions, the subject of crystal formation close to
half-filling for general values of $U/|J_p|,U/|J_h|$ has recently
been addressed using both numerical DMFT and analytical mean-field
analysis~\cite{DaoCapone}. The results are directly applicable to
loosely bound ALEs. Both DMFT and standard mean-field calculations
show the ALEs to form a superfluid condensate for moderate hopping
imbalances, and to phase separate into checkerboard crystal and
superfluid below some critical, density-dependent value of $J_h/J_p$
(c.f. figure~1 in~\cite{DaoCapone}). For a parameter regime
analogous to the one considered here (large hopping imbalance,
loosely bound ALEs) analytical mean-field calculations additionally
predict the possibility of pure crystal phase at noncommensurate
densities (c.f. figure~2 in~\cite{DaoCapone}). As the DMFT
calculations do not yield such a phase, the authors question the
validity of this result. Whether ALE-interactions can thus suppress
superfluidity at large hopping imbalances sufficiently for the
formation of a stable noncommensurate crystal phase in higher
spatial dimensions is not fully resolved yet.

As a final comment, note that the Hamiltonian (\ref{ham}) with
$J_p\neq J_h$ used here to describe ALEs, has been studied
previously in 1D~\cite{CazalillaGiamarchi} using Luttinger liquid
theory. There, the authors showed that increasing hopping imbalance
can lower the value $K_{c}$, the exponent controlling the algebraic
decay of correlation functions. For $|x-y|\gg 1$, density-density
(DD) correlations decay as $(\langle n_x n_y\rangle-\langle
n_x\rangle\langle n_y\rangle)\sim \cos(k_f(x-y)))/|x-y|^{K_c}$, with
$n_x=c_x^{\dagger}c_x+d_x^{\dagger}d_x$, whereas
pairing-correlations decay as $\langle c^{\dagger}_x d^{\dagger}_x
d_y c_y\rangle \sim 1/|x-y|^{1/K_c}$~\cite{Giamarchi}. As $J_h/J_p$
decreases, $K_c$ can decrease below $1$. Then crystaline ordering
becomes dominant over superfluid ordering, which is dominant for
$K_c>1$. The value of $J_h/J_p$ where $K_c=1$ will generally depend
on the exciton density $n_{ex}$. Based on the value of $K_c$ a phase
diagram can be drawn, c.f. figure~1 in \cite{CazalillaGiamarchi}.

\subsection{Phases of tightly bound ALEs}\label{ssection:tightALEreview}

As ALEs map onto an attractive Hubbard model with imbalanced hopping
of particles and holes, there are also some previously known results
for case of isotropic hopping $J_p^1=J_p^2=J_p^3=J_p$, yielding the
phase diagram in the tightly bound limit $|J_p|/U,|J_h|/U\ll 1$.
Here ALEs can be described as hard-core bosons with hopping rate
$J_{eff}$ and repulsive nearest-neighbour interactions $V_{eff}$.
For a regular square or cubic lattice, mean-field calculations (c.f.
e.g. \cite{MicnasRobaszkiewick}) predict either a checkerboard
'crystaline' Charge-Density-Wave (CDW) or a superfluid order at
maximal density $n_{ex}=\frac{1}{2}$, depending on the value of
$V_{eff}/J_{eff}$. The CDW is characterised by (DLRO), as was the
exciton crystal studied in the previous section. Away from maximal
filling, $n_{ex}<\frac{1}{2}$, mean-field calculations predict a
supersolid phase - the coexistence of superfluid and crystaline
order - for a broad range of $n_{ex}$ (c.f. figure~4
in~\cite{MicnasRobaszkiewick}). Beyond this range of densities, pure
superfluidity is predicted. However, Quantum Monte-Carlo simulations
\cite{BatrouniScalettar,HebertDorneich,KohnoTakahashi,KuklovSvistunov}
show the prediction of a supersolid away from commensurability to be
inaccurate for regular lattices. Depending on $V_{eff}/J_{eff}$ and
$n_{ex}$, tightly bound ALEs are either superfluid or undergo phase
separation into spatially disjoint subsystems, one exhibiting
superfluidity, the other checkerboard crystaline order \footnote{On
triangular - and thus frustrated - lattices, Monte-Carlo simulations
\cite{WesselTroyer} have shown that a supersolid phase can appear
however.}. Both Dynamical Mean Field Theory (DMFT) and standard
mean-field calculations suggest~\cite{DaoCapone} that when $J_p=J_h$
(i.e. $V_{eff}/J_{eff}=1$) tightly bound ALEs will be superfluid at
any density. One way to increase the value of
$V_{eff}/J_{eff}=(J_p^2+J_h^2)/(2J_pJ_h)$ and thereby attain phase
separation beyond some critical $n_{ex}$ is to have $J_p\neq J_h$
(c.f. \cite{Lee}). The realization of ALEs proposed here will always
have a significant imbalance, as the conduction band hopping rate
typically surpasses that of valence band by an order of magnitude.
Another possibility to increase the ratio $V_{eff}/J_{eff}$ is to
introduce additional nearest-neighbour interactions in the
Hamiltonian (\ref{ham}) (see~\cite{MicnasRobaszkiewick}), e.g. by
considering fermionic dipolar molecules instead of atoms.

In summary, while for sufficiently low $n_{ex}$, tightly bound
excitons always exhibit just superfluid order, exciton-exciton
interactions can enhance density-density correlations. For larger
values of $n_{ex}$ and with hopping imbalance between particles and
holes, they can result in phase-separation of the system into a
checkerboard ALE-crystal and an ALE superfluid.
\begin{figure}[htp]
\includegraphics[width=8.5cm]{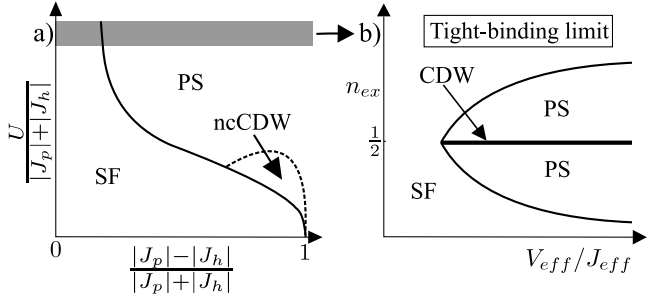}
\caption{(a) Sketch of ALE phase diagram as a function of hopping
imbalance and interaction strength relative to kinetic energy(after
figure~2 in ~\cite{DaoCapone}). (b) Sketch of phase diagram for
tightly bound ALEs at fixed density. SF: Superfluid, PS: Phase
separation, CDW: checkerboard Charge-Density Wave, ncCDW:
Charge-Density Wave at noncommensurate density. (a) Mean-field phase
diagram for ALEs at fixed density close to $n_{ex}=\frac{1}{2}$.
When the hopping imbalance $|J_p|-|J_h|/|J_p|+|J_h|$ is large
enough, the superfluid phase-separates into superfluid and
checkerboard-CDW. The separating line shifts with density. A pure
non-checkerboard CDW-phase might also be possible in a small
parameter range. For large $U/(|J_p|+|J_h|)$, ALEs become hard-core
bosons with ranged interactions. (b) Sketch of a qualitative phase
diagram for tightly bound ALEs. Doped away from the checkerboard CDW
phase at $n_{ex}=\frac{1}{2}$, ALEs phase-separate into superfluid
and checkerboard CDW, if $V_{eff}/J_{eff}$ is large enough. This
ratio can be tuned by changing $J_h/J_p$ or by introducing
longer-ranged interactions between particles and holes. }
\label{fig:phasediagram1}
\end{figure}
\section{Probing atomic lattice excitons}\label{section:spec}

Optical Lattice experiments provide a range of measurement
possibilities that we can use to determine the characteristic
properties of the many-body states that are produced. Three
important examples of this for probing ALEs are Lattice modulation
spectroscopy, RF Spectroscopy, and noise correlation measurements.

\subsection{Exciton detection via lattice modulation}
Periodic modulation of the lattice depth has become a standard way
to probe the excitation spectrum of many-body systems on an optical
lattice~\cite{SchoriEsslinger}. It is usually applied to systems in
the groundstate. If the frequency of the modulation matches that of
an excitation of the system, it drives transitions to higher-energy
states. Such a transition is then detected by ramping down the
confining potentials/interactions adiabatically, thereby
transferring all potential energy into kinetic energy of the
expanding atom cloud, which is then measured. This method has also
been used to drive transition from excited states to lower energy
states recently, in order to determine the energy of pairs of atoms
bound together by on-site repulsion~\cite{WinklerZoller}.

Qualitatively, probing for excitons via lattice modulations should
show a pronounced decrease in the systems total energy around a
frequency given by the bandgap minus the binding energy of the
exciton.

\subsection{Measuring the exciton condensate fraction via RF-spectroscopy}\label{ssection:rfspec}
\begin{figure}[htp]
\includegraphics[width=8.5cm]{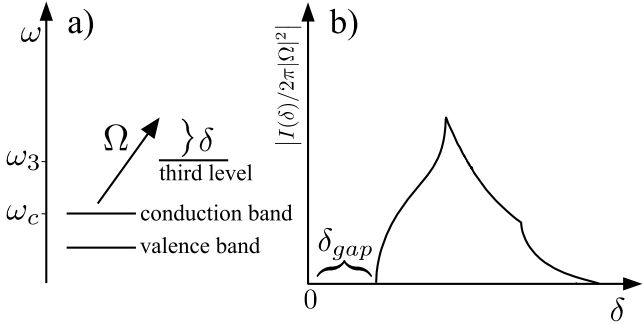}
\caption{RF spectroscopy for an ALE condensate. (a) Excitation with
RF-coupling $\Omega$ and detuning $\delta$ between conduction band
atoms and a third internal state. (b) Transfer rate
$|I(\delta)/2\pi\Omega^2|$ over detuning $\delta$ for $U/|J_p|=-8$,
$U/|J_h|=-40$, $n_{ex}=0.3$. The gap in the spectrum $\delta_{gap}$
is given by $\varepsilon_{0}^{(3)}+\min_{{\bf k}} E^{(2)}_{{\bf
k}}-\mu_c$. The kinks result from van Hove singularities.}
\label{fig:rfspec}
\end{figure}

In order to demonstrate particle-hole pairing and in particular to
detect the macroscopic occupation of the exciton groundstate, it is
possible to employ a range of measuring techniques that have been
developed for cold atoms experiments. At zero temperature, with no
thermal excitation of collective exciton modes with
$\textbf{K}\neq\textbf{w}$, the pairing amplitude $\Delta^P$ is
directly proportional to the condensate fraction. It can be probed
by the application of RF-spectroscopy, following the example of
\cite{TormaZoller} (c.f.~\cite{ChinGrimm} for an experimental
application). An RF-pulse with frequency $\omega_{RF}$ is used to
couple the conduction band atoms, at energy $\omega_c$, to another
internal state with energy $\omega_3$, with a detuning
$\delta=\omega_{RF}-(\omega_3-\omega_c)$, c.f.
figure~\ref{fig:rfspec}. Choosing $\delta>0$, the particle-hole
pairs are broken up in this process. It is assumed that this state
is still lattice-confined, with single particle energy
$\xi^{(3)}_{{\bf k}}=\varepsilon^{(3)}_{{\bf k}}-\mu_{3,eff}$, where
$\varepsilon^{(3)}_{{\bf k}}$ is the dispersion relation of the
state on the lattice and $\mu_{3,eff}$ is the effective chemical
potential, i.e. the chemical potential including any mean-field
shift from interactions.

The RF-pulse with frequency $\omega_{RF}$ has an amplitude $\Omega$
that can be assumed to be slowly varying on length scales of the
lattice. The transfer rate into the third state as a function of the
RF-detuning, $I(\delta)$, can be calculated by generalizing the
results in~\cite{TormaZoller} to the case of ALEs with anisotropic
conduction band hopping $J_p^d$ (and therefore $E_{\bf k}^{(1)}\neq
E_{\bf k}^{(2)}$). The result (at zero temperature) then becomes
\begin{equation}\label{spec}I(\delta)=-2\pi|\Omega|^2\sum_{{\bf k}}v_{{\bf k}}^2\delta(\xi^{(3)}_{{\bf k}}+\tilde{\Delta}+E^{(2)}_{{\bf k}})
\theta(\xi^{(3)}_{{\bf k}}).\end{equation} Here,
$\tilde{\Delta}=\mu_3-\mu_c-\delta$, and
$\delta=\omega_{RF}-(\omega_3-\omega_c)$ denotes the detuning of the
RF-pulse from the transition frequency between the two internal
states. $\mu_c$ is the chemical potential of the conduction band
atoms, for which relation $\mu_c+\mu_h=\mu_{ex}$ holds. This
expression is then valid generally for any $J_p$ and $J_h$, assuming
positive detuning.

The spectrum described by~(\ref{spec}) has a gap, given by the
minimum value of the $\delta$-function argument. Its value is
$\delta_{gap}=\varepsilon_{0}^{(3)}+\min_{{\bf k}} E^P_{{\bf
k}}-\mu_c$, assuming the additional internal state is initially
unpopulated and does not scatter off the other states, i.e
$\mu_{3,eff}=\mu_3=0$. If $n_{ex}$ is known,  $\Delta^P$ can be
determined selfconsistently from $\delta_{gap}$ using
(\ref{aledisp}).

\subsection{Detecting crystal structure and pairing correlations via noise spectroscopy}\label{ssection:noisespec}
We can detect both crystal structure and pairing correlation of ALEs
using atom shot-noise measurements, a technique that was proposed
in~\cite{AltmanLukin} and has been demonstrated in the
laboratory~\cite{FollingBloch,GreinerJin}. The former can be
identified through second-order correlation functions of
valence-band atoms, and the latter from second-order correlation
functions of atoms in two bands. Each of these can be obtained from
fluctuations in the density profile of the atomic gas.

For the ALE crystal, we expect holes to be anticorrelated with each
other around the Fermi-edge. As the operators
$b_{\textbf{k}}^{\dagger}$ for valence-band atoms in are related to
the hole operators through
$b_{\textbf{k}}^{\dagger}=d_{-\textbf{k}}$, we thus likewise expect
an anticorrelation peak around the Fermi-edge for the valence-band
atoms. In the case of the ALE-condensate, which is characterised by
the pairing of particles and holes at opposite quasimomenta,
pronouced anticorrelation of conduction- and valence-band atoms
should be visible at equal quasimomentum around the Fermi-edge, i.e.
"antipairing".

\begin{figure}[htp]
\includegraphics[width=8.5cm]{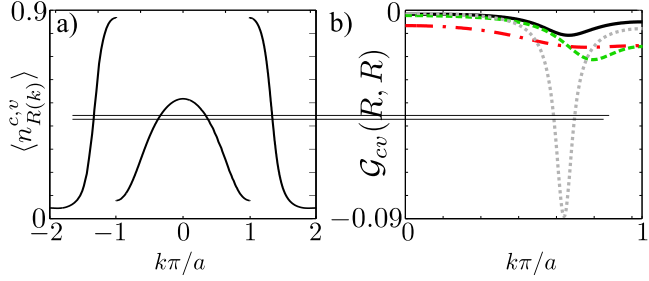}
\caption{(a) Density profile for quasimomentum states in the first
two Brillouin zones for 1D ALEs, calculated for a $40$-site lattice
in the crystal regime ($n_{ex}=\frac{13}{40}$, $U/|J_p|=4$,
$J_h/J_p=0.025$). (b) Pairing anticorrelations at $k=k'$ between
conduction- and valence-band atoms in different parameter regimes on
a $40$-site lattice. With parameters such that we expect to observe
an ALE crystal (solid line, $n_{ex}=\frac{13}{40}$, $U/|J_p|=4$,
$J_h/J_p=0.025$), antipairing is strongly supressed as compared to a
system outside this regime (other lines), with a minimum at the
Fermi-edge, $n_{ex}\pi/a$ away from the minimum of the Bloch band at
$\pi/a$. For equal hopping (dotted line, $n_{ex}=\frac{13}{40}$,
$U/|J_p|=4$, $J_h/J_p=1$) antipairing is strong and dominant (see
figure~\ref{fig:1dnoisecor}). At lower densities (dashed line,
($n_{ex}=\frac{9}{40}$, $U/|J_p|=4$, $J_h/J_p=1$) antipairing is
still stronger than for the cystal, with the minimum shifted due to
lower density. For increased attraction (dash-dotted line,
$n_{ex}=\frac{13}{40}$, $U/|J_p|=10$, $J_h/J_p=1$) antipairing is
slightly increased over the crystal} \label{fig:1dddprof}
\end{figure}

To obtain the density profiles experimentally, and from them the
fluctuations, the Brillouin zones need to be imaged, i.e. lattice
quasimomentum ${\bf k}$ needs to be mapped to real space position
${\bf R}({\bf k})$ on the detector. This is achieved by ramping down
the lattice sufficiently slowly to keep the atoms within their
respective bands whilst preserving their quasimomentum
(c.f.~\cite{KohlEsslinger}). Valence-band atoms then occupy the
first, and conduction-band atoms the second Brillouin zone (c.f.
figure~\ref{fig:1dddprof}a and~\ref{fig:densprofile}a). If density
fluctuations are just limited by shot noise, density fluctuations of
conduction and valence band atoms will be correlated according to
the connected correlation function
\begin{equation}\label{gcv}
\mathcal{G}_{cv}({\bf R},{\bf R}')=\langle n^c_{{\bf R}({\bf k})}
n^v_{{\bf R}({\bf k}')}\rangle- \langle n^c_{{\bf R}({\bf k})}
\rangle \langle n^v_{{\bf R}({\bf k}')}\rangle,
\end{equation}
whereas the density-fluctuations of valence-band
atoms are correlated amongst themselves according to
\begin{equation}\label{gvv}
\mathcal{G}_{vv}({\bf R},{\bf R}')=\langle n^v_{{\bf R}({\bf k})}
n^v_{{\bf R}({\bf k}')}\rangle- \langle n^v_{{\bf R}({\bf k})}
\rangle \langle n^v_{{\bf R}({\bf k}')}\rangle .
\end{equation}
Here $n^c_{{\bf
R}({\bf k})}=c_{{\bf R}({\bf k})}^{\dagger}c_{{\bf R}({\bf k})}$ and
$n^v_{{\bf R}({\bf k}')}=b_{{\bf R}({\bf k}')}^{\dagger}b_{{\bf
R}({\bf k}')}$.

\begin{figure}[htp]
\includegraphics[width=8.0cm]{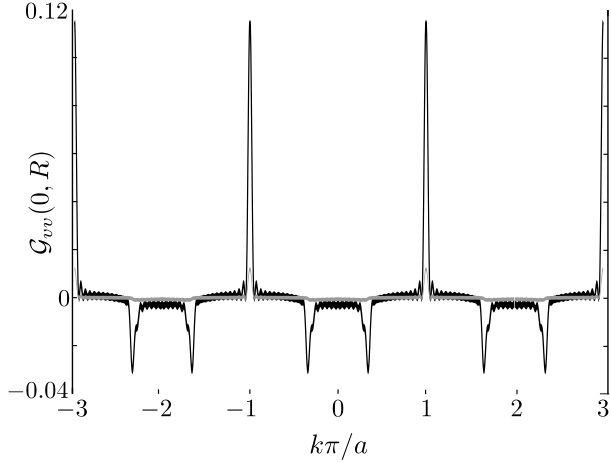}
\caption{Second order correlation function for valence band atoms,
computed on a $40$-site lattice. These represent results that can be
obtained from noise correlation measurements, in which the crystal
structure of ALEs appears. In the appropriate regime (black line,
$n_{ex}=\frac{13}{40}$, $U/|J_p|=4$, $J_h/J_p=0.025$), dips appear
at $\pm n_{ex}\pi/a$ away from the location of reciprocal lattice
vectors corresponding to the optical lattice. These signalize the
formation of the ALE crystal, as holes become localized through
atom-mediated long-range repulsions (see figure~\ref{fig:bopot}),
and occur at momentum values corresponding to the reciprocal lattice
vector of the exciton crystal. Such a result is representative of
the crystal regime. If we increase $U/|J_p|$ to $10$ (not shown),
the dip visibility decreases only slightly, and the crystal
structure still appears. Increasing $J_h/J_h$ to 1 (grey line)
causes the crystal and thus the dips to disappear, as does
decreasing $n_{ex}$ to $\frac{9}{40}$ (not shown)}
\label{fig:1dnoisecor}
\end{figure}

\begin{figure}[htp]
\includegraphics[width=8.5cm]{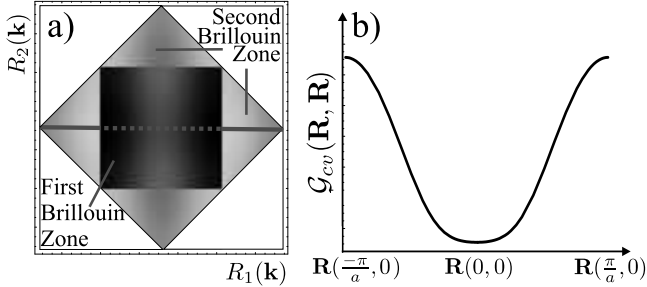}
\caption{(a) Calculated density profile of conduction and valence
band atoms as they would appear after ideal time-of-flight imaging.
Dark areas denote high atom density. The third dimension is
column-integrated. The first and second Brillouin zone are mapped to
real-space positions for the condensate ground state with
$n_{ex}=0.3$, $U/|J_p^1|=-8$, $J_p^2/J_p^1=J_p^3/J_p^1=-0.5$,
$U/|J_h|=-40$. (b) Calculated example of noise correlations between
conduction and valence band atoms, at positions ${\bf R}({\bf
k})={\bf R}({\bf k}')$. Here ${\bf R}({\bf k})$ is taken along the
dotted line in figure a, and ${\bf R}({\bf k}')$ is taken along the
solid line. The anticorrelation of conduction and valence band atoms
is most pronounced around the Fermi-edge of atoms and holes. As
$J_p^1<0$, $J_p^2,J_p^3>0$, the Fermi-edge in figure a lies along
the $k_2$ axis. The anticorrelation appears reduced because of
density-integration along $\textbf{R}(0,0,k_3)$.}
\label{fig:densprofile}
\end{figure}

For the case of ALEs in 1D we used the numerical algorithm employed
in section~\ref{ssection:1dcrystal} to compute both correlation
functions in different parameter regimes. These are plotted in Figs.
\ref{fig:1dddprof} and \ref{fig:1dnoisecor}, which when contrasted
display the crystal and superfluid characteristics of the system in
different parameter regimes. figure~\ref{fig:1dddprof}a shows the
momentum-space density profile for excitons in the first and second
Brillouin zones for the example of a parameter regime where we
expect a crystal to form. The fluctuations that would occur in the
experimental measurement of this density profile can be used to
compute the second order correlation functions (\ref{gcv}) and
(\ref{gvv})~\cite{AltmanLukin}. Figure~\ref{fig:1dddprof}b depicts
$\mathcal{G}_{cv}(R,R')$, that results from Brillouin-zone resolved
measurement, for different system parameters. It shows how
antipairing is suppressed when $n_{ex}$ increases and $U/|J_p|$
decreases. Figure ~\ref{fig:1dnoisecor} shows the periodic dips at
$G\pm n_{ex}\pi/a$ in $\mathcal{G}_{cv}(R,R')$ (where $G$ is any
reciprocal lattice vector). This can be measured from the noise
correlations measured from real momentum distributions, as is done
in  switching off the lattice suddenly, analogous to previous
experiments ~\cite{FollingBloch,GreinerJin}. The periodic
anti-correlations indicate the crystal, and their locations relative
to the large positive correlations reflect the reciprocal lattice
vector of the exciton crystal. Dips appear here because we are
computing noise correlations for fermions \cite{AltmanLukin}. Taken
together, the figures further illustrate the analysis of
section~\ref{ssection:1dcrystal}. They show that pairing and
density-density correlations are complementary, with pairing being
strongly suppressed and density-density correlations around the
Fermi-edge enhanced in the limit of strong hopping imbalance and
$n_{ex}\approx \frac{1}{3}$, $U/|J_p|=2$ $-$ $3$. This indicates
onset of the exciton crystal. Paring increases and the dips in the
density-density correlations vanish when $n_{ex}$ decreases or
$J_h/J_p$ increases, corresponding to the dominance of superfluid
character in the system.

For higher dimensional systems, the existence of a pure crystal
phase is not resolved, though crystal and superfluid order can
coexist (see sec.~\ref{ssection:tightALEreview}). When only
superfluid order is present, the correlations between conduction-
and valence-band atoms $\mathcal{G}_{cv}({\bf R},{\bf R}')$ can be
expressed through the mean-field coefficients $u$ and $v$:
\begin{equation}\mathcal{G}_{cv}({\bf R},{\bf R}')
=\left\{\begin{array}{cc}-u_{{\bf R}({\bf k})}^2v_{{\bf R}({\bf
k})}^2&{\bf R}({\bf k}) ={\bf R}({\bf k}')\\0&{\bf R}({\bf k})\neq
{\bf R}({\bf k}')\\\end{array}\right. .\end{equation} An example
plot of the column-integrated correlation function contained in the
shot noise is shown in figure~\ref{fig:densprofile}b.
Considering that the correlation at ${\bf k}={\bf k}'$ equals
$u_{{\bf k}}^2v_{{\bf k}}^2$ in the mean-field theory, this approach
provides an alternative for determining the condensate fraction via
the definition of $\Delta^P$.

\section{Exciton Formation in an Optical Lattice}\label{section:formation}
\subsection{General remarks}

One method to prepare excitons in an optical lattice would be to
create a band insulator of spin polarised fermions in the valence
band, and then to excite atoms to the conduction band, using, e.g.,
a Raman transition. This would be most directly analogous to the
method by which excitons are produced in the context of solid state
systems.

However, in the context of Atomic lattice excitions, it would be
desirable both to have control over the number of excitons produced,
and to have a means of producing these excitons in the lowest
possible many-body energy state. In contrast to excitons in the
solid state, the weak coupling of Atomic lattice excitions to
dissipative processes means that excitons prepared with a
significant  energy above that of the groundstate will not be
naturally cooled to this state. As a consequence, particle-hole
pairs cannot be created in an arbitrary manner if the ground state
is the desired outcome.

\subsection{A preparation scheme}\label{ssection:melt}
One scheme that we have investigated in detail for its efficiency in
preparing a many-exciton state close to the condensate groundstate
is a melting scheme adapted from~\cite{RablZoller}. Through
adiabatic tuning of the system parameters, a chosen filling factor
of ALEs can be prepared at essentially $T=0$ (this is in the same
spirit as other adiabatic preparation proposals, such as
\cite{TrebstZoller}).

\begin{figure}[htp]
\includegraphics[width=8.5cm]{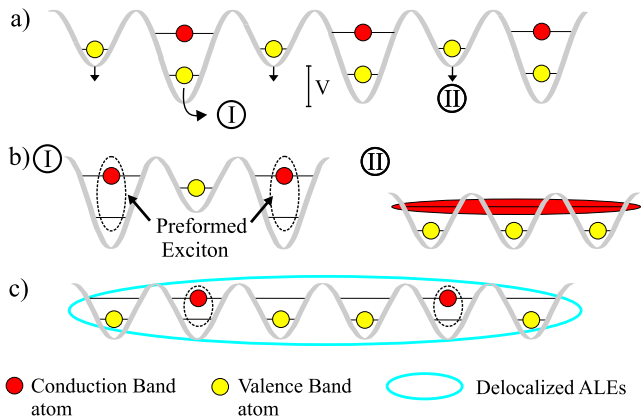}
\caption{Melting scheme to obtain an ALE condensate. (a) Initially
the ground state is formed in the presence of a superlattice, giving
site offsets $V$. Two alternative melting schemes, I and II, proceed
from there. I: (b) ALEs are preformed locally by removal of the atom
in the lower motional state on doubly occupied sites. c) Melting the
state by ramping $V$ down adiabatically yields delocalised ALEs. II)
(b) Groundstate of delocalised fermions in the upper band is formed,
melting the initial state by adiabatically decreasing $V$.
(Depending on sign of the hopping for ALEs, this must be done in the
lower band instead, see text for details) (c) The resulting state is
frozen by ramping up the lattice suddenly, and atoms in the lowest
band are removed from doubly occupied sites. This automatically
pairs each atom in the upper band with a hole in the lower band, in
the localised state expected in the limit $U/J\rightarrow \infty$.
Ramping the lattice down adiabatically then yields ALEs close to the
ground state.}\label{fig:melting}
\end{figure}

We begin with a spin-polarised Fermi gas loaded into an optical
lattice with an applied superlattice, which gives certain lattice
sites an offset energy $V$. The initial configuration of atoms is
loaded so that  the lowest motional state in each lattice site is
occupied, as well as the first excited state in wells with lower
energy, as shown in figure~\ref{fig:melting}a. This can be achieved,
e.g., using the methods discussed
in~\cite{RablZoller,ViveritSmerzi,GriessnerZoller}. The final
filling factor is chosen by the periodicity of the superlattice with
respect to the basic lattice period. In each case, we will want to
generate at some point an interaction energy $U$ between atoms in
the first excited motional level and atoms in the ground motional
level.

This can be achieved, e.g., by introducing an off-resonant Raman
coupling between the initial internal state of the spin-polarised
fermions and a second internal state. If the detuning is chosen so
that atoms in the excited motional level in the initial internal
state are coupled to the second internal state near to resonance
with the first motional level, then the coupling will be much
further detuned for the atoms in the initial internal state in the
lowest band than for those in the upper band. This will produce a
different admixture of the second internal state for atoms in each
band, and the non-identical internal states will allow s-wave
interactions between atoms in different bands.

A simple scheme for the adiabatic formation of excitons is now the
following: (i) Atoms are removed from the lowest motional level in
sites containing two atoms, by transferring them to a different
internal state and removing them from the lattice (see
figure~\ref{fig:melting}b-I). This is possible because the resonant
energy of the transition to a different internal state is shifted by
differences in interaction energies, and because it is possible to
address different bands using spin-dependent lattices for the
initial and final states of the transfer~\cite{RablZoller}. (ii) The
interaction $U$ is ``switched on''. (iii) $V$ is adiabatically
decreased to zero, effectively melting the preformed excitons and
forming the desired exciton gas  (see figure~\ref{fig:melting}c).

Until the last step, the state is protected by a gap of order $V$,
but the last step must be performed adiabatically with respect to
the exciton tunnelling rate, which for $J_p,J_h \ll U$ can be
estimated as $J_{ex} \approx 2 J_p J_h/U$.

These timescales can be improved upon by slightly modifying the
above scheme. Instead of removing atoms initially in step (i), $V$
is adiabatically decreased to zero, forming the ground state for
delocalised fermions in the first Bloch band, as shown in
figure~\ref{fig:melting}b-II. If the lattice depth is then suddenly
increased to a large value where $J_p, J_h \approx 0$ and atoms then
removed as in (i), then we preform the ground state of hard-core
excitons. If we then switch on the interaction $U$ and adiabatically
decrease the lattice depth, we obtain the desired exciton ground
state. Note that in this scheme we must take care of the sign of the
effective hopping for excitons, which in some cases requires an
additional complication. The atoms that we delocalise in the first
step should have the same hopping rate as the composite exciton.
Thus, for $J_p,$ $J_h<0$, and $U>0$, we obtain $J_{ex}>0$, and
therefore atoms should first be delocalised in the lowest band, then
transferred to the upper band whilst the lattice is very deep. This
requires atoms to be initially loaded in two internal states to
allow for double occupation in the lowest band.

\begin{figure}[htp]
\includegraphics[width=8.5cm]{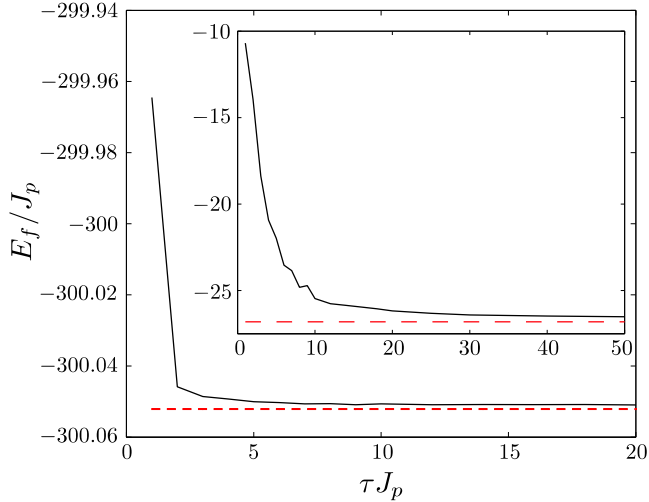}
\caption{Results of the Melting scheme II, as depicted in
figure~\protect{\ref{fig:melting}}. Main graph shows final state
energy $E_f/J_p$ obtained from numerical calculation of decreasing
$U/J$ for ALEs that were preformed utilizing the groundstate of
delocalised atoms (see figure~\protect{\ref{fig:melting}}b-II, c).
Parameters are: 60 sites, 15 ALEs, $U/|J_p|=U/|J_h|=20$. Exact
groundstate energy is indicated through a dashed red line. The inset
shows $E_f/J$ for the preparation of the groundstate of free atoms
in the upper band from the intial band insulator (see
figure~\ref{fig:melting}a, b-II), for identical parameters. Exact
groundstate energy is indicated by dashed red
line.}\label{fig:meltingresults}
\end{figure}

The advantages of the latter scheme are clear from
figure~\ref{fig:meltingresults}, where we show the energy of the
final state in 1D as a function of ramp speed in the final step. We
produce a final state with 15 excitons on 60 lattice sites, with
$U/|J_p|=U/|J_h|=20$, by ramping $V$ and then $J$ on a timescale
given in units of $J_{\rm final}^{-1}$. These results were computed
using a time-dependent DMRG
algorithm~\cite{Vidal2,Vidal1,DaleyVidal,WhiteFeiguin}, and show
clearly that ramping on a timescale much smaller than that implied
by $2 J_p J_h/U$ produce states with energies very near the ground
state (indicated by the dashed line in the figure). The main curve
shows the result for adiabatic ramping in the last step beginning
from the grounds state of hard core atoms in single atoms, and the
inset shows the energy of a single atom state after a similar ramp.

\section{Summary and Outlook}\label{section:summary}
Excitons are composite objects of fundamental interest in the
context of semiconductor physics. However, the theoretical models
presented for these systems relate even more closely to the
situation we discuss in optical lattices, making ALEs an important
testbed for study of the most interesting properties of these
objects. In particular, the availability of techniques for state
preparation and measurement in optical lattices provide tools to
study excitons that strongly completment those available in
semiconductor systems. ALEs can also be realised in parameter
regimes that strongly contrast with those available in
semiconductors. In this way we not only obtain the possibility for
condensation of these objects, but also for preparation of
additional phases, including an exciton crystal. Optical Lattices
also provide the possibility to directly reduce the dimensionality
of the system, making the important case of 1D excitons directly
accessible.

There are several open paths for the study of ALEs. Firstly, many of
the results we have discussed in this work, particularly pertaining
to the exciton crystal could be generalised to 2D and 3D. The
structure of degenerate excited bands \cite{LiuWu} could also be
introduced as an extra element in the higher dimensional context. In
addition, the study of composite objects could be developed in the
direction of multiple bound ALEs, in the case that, e.g., attractive
interactions are generated between excitons. Other possibilities
also arise to study systems analogous to excitonium~(see
\cite{MoskalenkoSnoke}, ch. 10), a dynamically created collection of
excitons that exists in the limit where the gap between valence of
conduction-band is small compared with the atom-atom interactions.

As a final comment, it is clear that there is a strongly connection
between ALEs and superconductivity or superfluidity of fermions via
the attractive Hubbard model, which we obtain here for interacting
particles and holes. It would also be possible to consider the case
of repulsive interactions between particles and holes, in which case
one could obtain repulsively bound ALEs, in the sense that the two
particle bound state appears above the scattering continuum. The
model describing this system would be a Hubbard model with repulsive
interations, which may lead to interesting analogies with d-wave
pairing states.

\ack{The authors acknowledge insightful discussions with H. P.
B\"uchler, M. Combescot,  and M. Szymanska. A. K. would like to
thank T. Giamarchi for helpful discussions on the hopping-imbalanced
Hubbard model and Luttinger liquids. Work in Innsbruck was supported
by Austrian Science Foundation (FWF), and by the European Commission
through the Integrated Project FET/QIPC SCALA. This work was
supported by Academy of Finland (project number 213362) and
conducted as part of a EURYI scheme award. See www.esf.org/euryi.}

\appendix
\section{Solution for the Single Exciton}\label{sec:app1}
As for the COM-motion,~(\ref{exeq}) with $U=0$ corresponds to a
Schr\"odinger equation for a free particle on a lattice. That
equation is resolved by the Green's function with discrete
translation invariance,
\begin{equation}G_0({r}-{r}')=a\int _{-\pi/a}^{\pi/a} \frac{\rmd k}{2\pi}\frac{\exp\left[\rmi k({r}-{r}')\right]}{E+2J\cos(ka)+\rmi \eta}.\end{equation}
For $U>0$ - the single particle motion with an impurity at
lattice-site ${r}=0$ - the Kroneker-delta in~(\ref{exeq}) allows an
easy, explicit resummation of the recursive relation for the full
Green's function:
\begin{equation}\Rightarrow G({r},{r}') = G_0({r}-{r}')+\frac{G_0({r})G_0({r}')}{1-G_0(0)/U}\end{equation}
One can read off from this expression that the full Green's function
has a single bound state pole for each value of $K$ with energy
$E_K$, where the denominator of the second term vanishes, i.e., for
$U^{-1}=G_0(0)$. The numerator yields directly the unnormalised
bound state wavefunction in relative coordinates,
$\rho_{{r}}=CG_0({r})$, which is given by
\begin{equation}\rho_{{r}}=Ca\int_{-\pi/a}^{\pi/a} \frac{\rmd k}{2\pi}
\frac{\exp\left[\rmi kr\right]}{E+ 2J\cos(ka)}\end{equation} where
$C$ is the normalisation constant. This is depicted in
figure~\ref{fig:exdisp} a,b. The dispersion relation $E$ of the
exciton, which forms the exciton solution in figure~\ref{fig:exdisp}
is determined by the implicit equation
\begin{equation}\label{ek}U^{-1}=a\int_{-\pi/a}^{\pi/a} \frac{\rmd k}{2\pi}\frac{1}{E+ 2J\cos(ka)},\end{equation}
which can be integrated to give the result quoted in section
\ref{ssection:singale}.

\section{Solution to the $N$-impurities problem}
\label{app:Nimpurities} In solving the Schr\"odinger equation for
ALEs in the Born-Oppenheimer (BO) approximation, we must treat the
problem of $N$ spin-polarised fermions on a lattice (the conduction
band atoms), moving in a potential provided by the $N$ holes, which
essentially act as impurities. The BO Hamiltonian
from~(\ref{HBOequation}) can be decomposed into a sum of single-atom
Hamiltonians $h_n[\underline{\textbf{R}}]$. As atom-hole interaction
is on-site, we can write
\begin{equation}\label{HBOdecompse}H_{BO}[\underline{\textbf{R}}]=\sum_{n=1}^N
h_n[\underline{\textbf{R}}],\quad
h_n[\underline{\textbf{R}}]=-J_p\widetilde{\Delta}_{\textbf{r}_n}-U\sum_{n'=1}^N\delta_{\textbf{r}_n\textbf{R}_{n'}}\end{equation}
Each of the single-atom Hamiltonians $h[\underline{\textbf{R}}]$ has
eigenstates $\rho_k(\textbf{r};\underline{\textbf{R}})$, obtained
from the solving the eigenvalue equations
\begin{equation}\label{BOspeigeneq}h[\underline{\textbf{R}}]\rho_k(\textbf{r};\underline{\textbf{R}})=E_k[\underline{\textbf{R}}]\rho_k(\textbf{r};\underline{\textbf{R}})\end{equation}
where $k$ runs over all bound and scattering solutions. In general,
depending on dimensionality and the ratio of $U/J_p$, there will be
up to $N$ bound state solutions for $N$ holes for any configuration
$\underline{\textbf{R}}$. If the impurities approach each other too
closely (on a length scale dependent on $U/|J_p|$), some of the
bound states will disappear.  In the following, we assume that the
system is sufficiently dilute, that $N$ bound states will always
exist. For $N=2$ this will always be the case provided
$U/|J_p|\gtrsim 2$.

From the $\rho_k(\textbf{r};\underline{\textbf{R}})$'s, the exact
many-atom wavefunctions
$\phi_{\alpha}(\underline{\textbf{r}};\underline{\textbf{R}})$,  can
be constructed by forming the Slater determinant:
\begin{equation}\label{BOslater}\phi_{\alpha}(\underline{\textbf{r}};\underline{\textbf{R}})
=\left|\begin{array}{ccc}
\rho_{k_1}(\textbf{r}_1;\underline{\textbf{R}}) & \ldots & \rho_{k_1}(\textbf{r}_N;\underline{\textbf{R}})\\
\vdots & \ddots & \vdots\\
\rho_{k_N}(\textbf{r}_1;\underline{\textbf{R}}) & \ldots &
\rho_{k_N}(\textbf{r}_N;\underline{\textbf{R}})
\end{array}\right|\end{equation}
The wavefunction
$\phi_{\alpha}(\underline{\textbf{r}};\underline{\textbf{R}})$ is
antisymmetric in $\underline{\textbf{r}}$ and
solves~(\ref{BOeigeneq1}). The index $\alpha$ thus is a multiindex,
with $\alpha=(k_1,\ldots,k_N)$ and the condition $k_a\neq k_b,$ for
$a\neq b$, and the BO-energies $E_{\alpha;BO}$ of~(\ref{BOeigeneq1})
are obtained from summing the single-atom energies:
\begin{equation}E_{\alpha;BO}[\underline{\textbf{R}}]=\sum_{\{k_1,\ldots,k_N\}=\alpha}
E_{k_a}[\underline{\textbf{R}}]\end{equation}

We can then obtain the explicit form of the single-atom
wavefunctions $\rho_k(\textbf{r};\underline{\textbf{R}})$, i.e. the
solutions to~(\ref{BOspeigeneq}). As the single atom Green's
function $G(\textbf{r}',\textbf{r};\underline{\textbf{R}})$ has the
spectral decomposition
$G(\textbf{r}',\textbf{r};\underline{\textbf{R}})=\sum_k\rho_k^*(\textbf{r}';\underline{\textbf{R}})\rho_k(\textbf{r};\underline{\textbf{R}})/(E-E_k[\underline{\textbf{R}}]+\imath\eta)$,
(we do not write the $E$-dependence of the Green's function
explicitly in the following), knowing
$G(\textbf{R}_n,\textbf{r};\underline{\textbf{R}})$ gives access to
both eigenenergies $E_k[\underline{\textbf{R}}]$ and wavefunctions
(from poles and residues respectively).

As the full single atom wavefunction obeys
\begin{equation}G(\textbf{r}',\textbf{r};\underline{\textbf{R}})=G_0(\textbf{r}',\textbf{r})+\sum_{\textbf{r}'}G_0(\textbf{r}',\textbf{r}')V(\textbf{r}')G(\textbf{r}',\textbf{r};\underline{\textbf{R}})\end{equation}
where $V(\textbf{r}')=-U\sum_{n=1}^N
\delta_{\textbf{r}'\textbf{R}_n}$ and
\begin{equation}G_0(\textbf{r}',\textbf{r})=a^D\int_{B.Z.} \frac{\rmd^D k}{(2\pi)^D}\frac{\exp\left[\rmi \textbf{k}(\textbf{r}'-\textbf{r})\right]}{E+2J_p\sum_d\cos(k_da)+\rmi \eta}.\end{equation}
The set of $G(\textbf{R}_n,\textbf{r};\underline{\textbf{R}})$'s can
be determined from the linear system of equations
\begin{equation}\label{BOspLSE}\textbf{A}[E,\underline{\textbf{R}}]\textbf{G}(\textbf{r};\underline{\textbf{R}})=\textbf{G}_0(\textbf{r}),\end{equation}
where
\begin{equation}(\textbf{A}[E,\underline{\textbf{R}}])_{nm}=\left\{\begin{array}{cc}
1+UG_0(\textbf{R}_m,\textbf{R}_m)& m=n\\
UG_0(\textbf{R}_n,\textbf{R}_m)& m\neq n
\end{array}\right.\end{equation}
and
$\textbf{G}(\textbf{r};\underline{\textbf{R}})=(G(\textbf{R}_1,\textbf{r};\underline{\textbf{R}}),\ldots,G(\textbf{R}_N,\textbf{r});\underline{\textbf{R}})$,
$\textbf{G}_0(\textbf{r})=(G_0(\textbf{R}_1,\textbf{r}),\ldots,G_0(\textbf{R}_N,\textbf{r}))$.
The eigenenergies are determined from the condition
\begin{equation}\det(\textbf{A}[E_k[\underline{\textbf{R}}],\underline{\textbf{R}}])=0\end{equation}
and we assume here that this has $N$ bound-state solutions. We thus
see immediately that any $N$-hole single-atom bound state
wavefunction can always be written as a linear combination of
functions that have the form of single-hole bound states.

For example, for $N=2$~(\ref{BOspLSE}) is solved by
\begin{eqnarray}\fl G(\textbf{R}_1,\textbf{r};\underline{\textbf{R}})&=&\frac{(1+UG_0(\textbf{R}_1,\textbf{R}_1))G_0(\textbf{R}_1,\textbf{r})-UG_0(\textbf{R}_1,\textbf{R}_2)G_0(\textbf{R}_2,\textbf{r})}{(1+UG_0(\textbf{0},\textbf{0}))^2-U^2G_0^2(\textbf{R}_1,\textbf{R}_2)}\\
\fl
G(\textbf{R}_2,\textbf{r};\underline{\textbf{R}})&=&\frac{(1+UG_0(\textbf{R}_2,\textbf{R}_2))G_0(\textbf{R}_2,\textbf{r})-UG_0(\textbf{R}_1,\textbf{R}_2)G_0(\textbf{R}_1,\textbf{r})}{(1+UG_0(\textbf{0},\textbf{0}))^2-U^2G_0^2(\textbf{R}_1,\textbf{R}_2)}
\end{eqnarray}
where the eigenenergies $E_{\pm}[\textbf{R}_1,\textbf{R}_2]$ are
implicitly determined from the denominator being zero,
\begin{equation}(1+UG_0(\textbf{0},\textbf{0}))^2=U^2G_0^2(\textbf{R}_1,\textbf{R}_2),\end{equation}
and the residue yields two wavefunctions,
\begin{equation}\rho_{\pm}(r;\underline{\textbf{R}})=C(G_0(R_1,\textbf{r})\pm G_0(R_2,\textbf{r})),\end{equation}
where $C$ carries the normalization.

The lowest energy BO-potential is given by
$E_{+}[\textbf{R}_1,\textbf{R}_2]+E_{-}[\textbf{R}_1,\textbf{R}_2]$,
examples of which are plotted in figure~\ref{fig:bopot} for the 1D
case.

\section{Description of a many-exciton condensate}\label{sec:appA}
The condensed state of the excitons, in which a macroscopic number
of them occupies the single-exciton groundstate, breaks the symmetry
of the Hamiltonian associated with number conservation. Macroscopic
occupation of the groundstate $A_0^{\dagger}=\sum_kA_{0,0}({\bf
k},-{\bf k})c_{{\bf k}}^{\dagger}d_{-{\bf k}}^{\dagger}|0\rangle$
implies that $\langle c_{{\bf k}}^{\dagger}d_{-{\bf
k}}^{\dagger}\rangle\neq0$. A way to introduce symmetry breaking
into the Hamiltonian is by applying a Bogoliubov canonical
transformation to the particle and hole operators
\cite{KeldyshKozlov,MoskalenkoSnoke}:

\begin{eqnarray}\label{newop}
C_{{\bf k}}=\mathcal{D}c_{{\bf k}}\mathcal{D}^{\dagger}&=&u_{{\bf k}} c_{{\bf k}}+v_{{\bf k}} d^{\dagger}_{-{\bf k}}\\
D_{{\bf k}}=\mathcal{D}d_{{\bf k}}\mathcal{D}^{\dagger}&=&u_{{\bf
k}} d_{{\bf k}}-v_{{\bf k}} c^{\dagger}_{-{\bf k}}\end{eqnarray}

with $u_{{\bf k}}^2+v_{{\bf k}}^2=1$ to maintain anti-commutation
relations. Here $C_{{\bf k}}$, $D_{{\bf k}}$ correspond to new
quasiparticle and quasihole operators, that have vanishing
occupation number in the condensate groundstate of the system,
\begin{equation}\label{constr2}\langle C_{{\bf k}}^{\dagger}C_{{\bf k}}\rangle=\langle D_{{\bf k}}^{\dagger}D_{{\bf k}}\rangle=0,\end{equation}
which automatically imposes $\langle c^{\dagger}_{{\bf k}}
d^{\dagger}_{-{\bf k}}\rangle\neq 0$. Inserting the inverse of
(\ref{newop}) into (\ref{scham}) and rearranging to obtain a
normal-ordered form again, the Hamiltonian decomposes into a sum of
three terms:
\begin{equation}H=H_{const}+H_{2}+H_{Int}\end{equation}
These terms have the following structure:
\newline

- $H_{const}$ is a constant containing no operators
\newline

- $H_2$ is given by
\begin{equation}\label{h2}H_2=\sum_{{\bf k}} E^{(1)}_{{\bf k}} C_{{\bf k}}^{\dagger}C_{{\bf k}}
+E^{(2)}_{{\bf k}} D_{{\bf k}}^{\dagger}D_{{\bf k}}-F_{{\bf k}}(C^{\dagger}_{{\bf k}} D^{\dagger}_{-{\bf k}}+D_{-{\bf k}}C_{{\bf k}})\end{equation}
with
\begin{eqnarray}\label{disp1}\fl E^{(1)}_{{\bf k}}=E^p_{{\bf k}} u_{{\bf k}}^2-E^h_{{\bf k}} v_{{\bf k}}^2+\sum_{{\bf l}}\left[
u_{{\bf k}}^2\left(V_0+V_0^p-V^p_{{\bf l}-{\bf k}}\right)+v_{{\bf
k}}^2\left(V^h_{{\bf l}-{\bf k}}-V_0-V_0^h\right)\right]v_{{\bf
l}}^2\nonumber\\-2u_{{\bf k}} v_{{\bf k}}\sum_{{\bf l}} V_{{\bf
l}-{\bf k}}u_{{\bf l}} v_{{\bf l}}\end{eqnarray}
\begin{eqnarray}\label{disp2}\fl E^{(2)}_{{\bf k}}=E^h_{{\bf k}} u_{{\bf k}}^2-E^p_{{\bf k}} v_{{\bf k}}^2+\sum_{{\bf l}}\left[
u_{{\bf k}}^2\left(V_0+V_0^h-V^h_{{\bf l}-{\bf k}}\right)+v_{{\bf k}}^2\left(V^p_{{\bf l}-{\bf k}}-V_0-V_0^p\right)\right]v_{{\bf l}}^2\nonumber\\
-2u_{{\bf k}} v_{{\bf k}}\sum_{{\bf l}} V_{{\bf l}-{\bf k}}u_{{\bf
l}} v_{{\bf l}}\end{eqnarray}
\begin{eqnarray}\fl F_{{\bf k}}=u_{{\bf k}} v_{{\bf k}}\left[E^p_{{\bf k}}+E^h_{{\bf k}}-\sum_{{\bf l}} \left(2V_0+V^p_0+V^h_0-
V^p_{{\bf l}-{\bf k}}-V^h_{{\bf l}-{\bf k}}\right)v_{{\bf
l}}^2\right]\nonumber\\+\left(u_{{\bf k}}^2-v_{{\bf
k}}^2\right)\sum_{{\bf l}} V_{{\bf l}-{\bf k}}u_{{\bf l}} v_{{\bf
l}}\end{eqnarray}

The last term in (\ref{h2}) corresponds to a process where
electron-hole pairs with total momentum equal to zero are being
spontaneously created from and annihilated into the condensate.
\newline

- $H_{Int}$ is the transformed interaction part of the Hamiltonian
and contains all possible quartic combinations of particle and hole
operators. Among other processes, these correspond to the creation
and annihilation of two pairs of quasiparticles and quasiholes with
total momentum zero.
\newline

$H$ re-expressed in the new operators can thus be unstable with
respect to the spontaneous creation and annihilation of free
quasiparticle-quasihole pairs. To rectify this, Keldysh and Kozlov
demand that $u_{{\bf k}}$ and $v_{{\bf k}}$ be chosen such that they
satisfy the constraint~\cite{KeldyshKozlov}
\begin{equation}\label{constr}\langle C^{\dagger}_{{\bf k}}D_{-{\bf k}}^{\dagger}\rangle=\langle C_{{\bf k}} D_{-{\bf k}}\rangle=0.\end{equation}
The average here is performed over the exact groundstate, and thus
$u_{{\bf k}}$ and $u_{{\bf k}}$ need to be determined from the full
transformed Hamiltonian, subject to (\ref{constr}), which is a
difficult task.

It is simpler to satisfy constraint (\ref{constr}) just for the
groundstate of $H_2$ and disregard $H_{Int}$ for the moment. This is
equivalent to demanding that $F_{{\bf k}}=0$ for all ${\bf k}$
\cite{MoskalenkoSnoke}, which in turn is equivalent to $u_{{\bf k}}$
and $v_{{\bf k}}$ satisfying mean-field~(\ref{bcseq}) -
(\ref{bcsxi}). In this approximation, only particle-hole pairs at
opposite momenta are correlated. The Bogoliubov transformation that
generates these correlations is
\begin{equation}\mathcal{D}=\exp\left[\sqrt{n_{ex}}\sum_{{\bf k}} A({\bf k})(c^{\dagger}_{{\bf k}}d_{-{\bf k}}^{\dagger}-d_{-{\bf k}}c_{{\bf k}})\right]\end{equation}
where $u_{{\bf k}}=\cos(\sqrt{n_{ex}}A({\bf k}))$, $v_{{\bf
k}}=\sin(\sqrt{n_{ex}}A({\bf k}))$. When applied to the vacuum
particle-hole vacuum $|0\rangle$, $\mathcal{D}$ generates the
coherent state to the operators $c^{\dagger}_{{\bf
k}}d^{\dagger}_{-{\bf k}}$, $d_{-{\bf k}}c_{{\bf k}}$ which
automatically satisfies (\ref{constr2}) and (\ref{constr}). The new
groundstate is consequently given by
\begin{equation}\label{psi}|\psi\rangle=\mathcal{D}|0\rangle=\prod_{{\bf k}}(u_{{\bf k}}+v_{{\bf k}} c_{{\bf k}}^{\dagger}d_{-{\bf k}}^{\dagger})|0\rangle,\end{equation}
of the same form as the BCS-groundstate. The dispersion relations of
the single-quasiparticle excitations above this groundstate are
given by (\ref{disp1}) and (\ref{disp2}).

\end{document}